\documentclass[useAMS,usenatbib]{mn2e}

\usepackage{psfig,epsfig}
\usepackage{mathrsfs}
\usepackage{amssymb}
\usepackage{epsfig}
\usepackage{marvosym}
\usepackage{subfigure}
\usepackage{fancyhdr}
\usepackage{rotating}

\def\degr{\hbox{$^\circ$}}

\def\gsim{\mathrel{\hbox{\rlap{\lower.55ex \hbox {$\sim$}}
                   \kern-.3em \raise.4ex \hbox{$>$}}}}
\def\lsim{\mathrel{\hbox{\rlap{\lower.55ex \hbox {$\sim$}}
                   \kern-.3em \raise.4ex \hbox{$<$}}}}

\def\he{\hbox{He\,{\sc i} $\lambda$5875}}

\def\EGAPS{\hbox{\sl EGAPS\ }}
\def\UVEX{\hbox{\sl UVEX\ }}
\def\IPHAS{\hbox{\sl IPHAS\ }}
\def\WISE{\hbox{\sl WISE\ }}
\def\UKIDSS{\hbox{\sl UKIDSS\ }}
\def\SDSS{\hbox{\sl SDSS\ }}
\def\MASS{\hbox{\sl MASS\ }}

\def\RL1{\hbox{{$R_{L_{1}}$}}}

\title[UV-excess sources with a red/IR-counterpart: low-mass companions, debris disks and QSO selection.]
{UV-excess sources with a red/IR-counterpart: low-mass companions, debris disks and QSO selection}
\author[Kars Verbeek et al.]{{Kars Verbeek$^{1}$\thanks{E-mail:k.verbeek@astro.ru.nl}, 
Paul J. Groot$^{1}$,
Simone Scaringi$^{1,2}$,
Jorge Casares$^{3}$,}
\newauthor{
Jesus M. Corral-Santana$^{3,4}$
Niall Deacon$^{5}$
Janet E. Drew$^{6}$,
Boris T. G{\"a}nsicke$^{7}$,}
\newauthor{
Eduardo Gonz{\'a}lez-Solares$^{8}$,
Robert Greimel$^{9}$,
Ulrich Heber$^{10}$,
Ralf Napiwotzki$^{6}$,}
\newauthor{
Roy H. {\O}stensen$^{2}$,
Danny Steeghs$^{7}$,
Nicholas J. Wright$^{6}$
and Albert Zijlstra$^{11}$}\\
$^{1}$Department of Astrophysics, Radboud University Nijmegen,
  P.O. Box 9010, 6500 GL Nijmegen, The Netherlands\\
$^{2}$Instituut voor Sterrenkunde, KU Leuven, Celestijnenlaan 200D, B-3001 
  Leuven, Belgium\\
$^{3}$Instituto de Astrof\'{\i}sica de Canarias, Via Lactea, s/n
  E-38205 La Laguna (Tenerife), Spain\\
$^{4}$Departamento de Astrof\'{\i}sica, Universidad de La Laguna, 
  La Laguna E-38205, S/C de Tenerife, Spain\\
$^{5}$Max-Planck-Institute f{\"u}r Astronomie, K{\"o}nigstuhl 17, 69117, 
  Heidelberg, Germany\\
$^{6}$Centre for Astronomy Research, Science \& Technology Research
  Institute, University of Hertfordshire, Hatfield, AL10 9AB, UK\\
$^{7}$Physics Department, University of Warwick, Coventry, CV4 7AL,
  UK\\
$^{8}$Cambridge Astronomy Survey Unit, Institute of Astronomy, University of
  Cambridge, Madingley Road, Cambridge, CB3 0HA, UK\\
$^{9}$Institut f\"ur Physik, Karl-Franzen Universit\"at Graz,
Universit\"atsplatz 5, 8010 Graz, Austria\\
$^{10}$Dr. Remeis-Sternwarte Bamberg, Universit\"at Erlangen-N\"urnberg,
  Sternwartstrasse 7, D-96049 Bamberg, Germany\\
$^{11}$ Jodrell Bank Centre for Astrophysics, Alan Turing Building, 
  University of Manchester, M13 9PL, UK
}

\begin{document}

\date{Accepted for publication in MNRAS}

\pagerange{\pageref{firstpage}--\pageref{lastpage}} \pubyear{2013}

\maketitle

\label{firstpage}

\begin{abstract}
We present the result of the cross-matching between UV-excess sources 
selected from the UV-excess survey of the Northern Galactic Plane (\UVEX) 
and several infrared surveys (2\MASS, \UKIDSS and \WISE).
From the position in the $(J-H)$ vs. $(H-K)$ colour-colour diagram we select
UV-excess candidate white dwarfs with an M-dwarf type companion, 
candidates that might have a lower mass, brown-dwarf type companion, and candidates showing an 
infrared-excess only in the K-band, which might be due to a debris disk.
Grids of reddened DA+dM and sdO+MS/sdB+MS model spectra are fitted to the $U,g,r,i,z,J,H,K$ photometry
in order to determine spectral types and estimate temperatures and reddening.
From a sample of 964 hot candidate white dwarfs with $(g-r)<$0.2, 
the spectral energy distribution fitting shows that $\sim$2-4$\%$ of the white dwarfs have an M-dwarf companion, 
$\sim$2$\%$ have a lower-mass companion, and no clear candidates for having a debris disk are found.
Additionally, from \WISE 6 UV-excess sources are selected as candidate Quasi-Stellar Objects (QSOs). 
Two UV-excess sources have a \WISE IR-excess showing up only in the mid-IR $W3$ band of \WISE, 
making them candidate Luminous InfraRed Galaxies (LIRGs) or Sbc star-burst galaxies.\\
\end{abstract}


\newpage

\section{Introduction}
One of the main goals of the European Galactic Plane Surveys (\EGAPS) is to obtain a homogeneous sample 
of evolved objects in our Milky Way with well-known selection limits.
The \EGAPS data also contains more esoteric 
objects (e.g. Nova V458 Vul, Wesson et al., 2008; Necklace Nebula, Corradi et al., 2011; 
photo-evaporating prolyd-like objects, Wright et al., 2012).
Over the last years the data of large sky surveys yielded several
known white dwarfs with gas or dust disks (G{\"a}nsicke et al., 2007; G{\"a}nsicke et al., 2011; 
G{\"a}nsicke et al., 2008; Zuckerman and Becklin, 1987; Brinkworth et al., 2009; Brinkworth et al., 2012; Kilic et al., 2012; Debes et al., 2011).
When optical surveys are cross-matched with the data of infrared (IR) surveys 
(e.g. \SDSS-\UKIDSS: Silvestri et al., 2006; Heller et al., 2009; Girven et al., 2011, 
\SDSS-\WISE: Debes et al., 2011; Hoard et al., 2011, \SDSS-2\MASS: Hoard et al., 2007 and \IPHAS-2\MASS: Wright et al., 2008), 
the classification of sources can be extended, and hot white dwarfs with companions or 
debris disks, and other peculiar objects are detected.
The dusty debris disks around white dwarfs are believed to form during the destruction of asteroids, the remnants of 
the planetary systems that orbited the star earlier in its evolution at the main-sequence. 
Emission lines in the spectra of the white dwarfs indicates that there might also be gaseous material present in these disks.
This might also clarify the spectra of metal-polluted white dwarfs (G{\"a}nsicke et al., 2012; Debes, Walsh \& Stark, 2012; 
Dufour et al., 2012; Farihi et al., 2012; Debes \& Sigurdsson, 2002; Jura, 2003). 
The time for metals to sink out of the atmosphere of the white dwarf is in the order of a few days for hot DA white dwarfs,
and up to 10$^6$ years for DB/DC white dwarfs (Table 4 to 6 of Koester, 2009; Koester \& Wilken, 2006),
indicating that accretion is ongoing for most objects.
An unknown fraction of the \UVEX white dwarfs will have an M-dwarf companion (Silvestri et al., 2006; Farihi, Becklin \& Zuckerman, 2005; Debes, 2011).
From the cross-matching between infrared and optical observations, about 0.2-2$\%$ of the white dwarfs 
are expected to have an IR-excess due to a brown dwarf type companion, and 0.3-4$\%$ of the white dwarfs 
are expected to be debris disk candidates (Girven et al., 2011; Steele et al., 2011; Debes et al., 2011; 
Barber et al., 2012; Farihi, Becklin \& Zuckerman, 2005).\\

A number of the optically selected UV-excess sources from \UVEX, as described in Verbeek et al. (2012a; hereafter V12a), 
will show a near-infrared (NIR) and mid-infrared (MIR) excess due to a low-mass companion 
or due to interstellar and/or circumstellar material.
The UV-excess catalogue of V12a consists of a mix of different populations, such as white dwarfs, 
interacting white dwarf binaries, subdwarfs of type O and B (sdO/sdB), emission line stars and QSOs. 
Due to the limited statistics and inhomogeneity, the fraction of optically selected white dwarfs with an IR-excess due to a low-mass companion is
very uncertain in the Galactic Plane (Hoard et al., 2011). The UV-excess catalogue of V12a offers a complete white dwarf sample
for this purpose, eventhough the sample also contains other populations. 
A distinction between white dwarfs with a companion or disk and e.g. Young Stellar Objects (YSOs), Be stars and Cataclysmic Variables 
can be made using the strength of the H$\alpha$ emission (Witham et al., 2008, Corradi et al., 2010, Barentsen et al., 2011). 
A fraction of the subdwarf stars and A-type stars in the UV-excess catalogue might show an 
IR-excess (Hales et al., 2009), and UV-excess sources can also have both an optical blueness and infrared-excess when they are 
non-stellar, e.g. QSOs (Roseboom et al., 2012, Xue-Bing et al., 2012, Wright et al., 2010).\\

The \UVEX survey images a 10$\times$185 degrees wide band (--5\degr$<$ $b$ $<$+5\degr) centered on the Galactic equator
in the $U,g,r$ and $\he$ bands down to $\sim 21^{st}-22^{nd}$ magnitude using the Wide Field Camera mounted 
on the Isaac Newton Telescope on La Palma (Groot et al., 2009).
From the first 211 square degrees of \UVEX data a catalogue of 2\,170 UV-excess sources
was selected in V12a. These UV-excess candidates were selected from the $(U-g)$ versus $(g-r)$ colour-colour diagram 
and $g$ versus $(U-g)$ and $g$ versus $(g-r)$ colour-magnitude diagrams by an automated field-to-field selection algorithm.
Less than $\sim$1$\%$ of these selected \UVEX sources are known in the literature. Spectroscopic follow-up of 132 UV-excess candidates 
selected from \UVEX, presented in Verbeek et al. (2012b; hereafter V12b), shows that most UV-excess candidates are indeed genuine 
UV-excess sources such as white dwarfs, subdwarfs and interacting white dwarf binaries.\\

In this work we present the IR photometry of the UV-excess sources in the \UVEX catalogues of V12a.
Our goals are i) to see what fraction of the hot white dwarfs have a companion (late MS or BD), 
ii) to see if we can use IR photometry to select non-white dwarfs from our UV-excess catalogue, and
iii) to see if we can find any debris disks.
In Sect.\ \ref{sec:crossmatching} the cross-matching of the full UV-excess catalogue with IR/red surveys is presented. 
In Sect.\ \ref{sec:WDIR} hot UV-excess candidate white dwarfs with $(g-r)$$<$0.2 with an IR-excess are selected 
and classified by fitting grids of reddened DA+dM and sdO/sdB models to the optical and infrared photometry.
The spectral types of companions later than M6 are determined from the IR-excess.
From these results the fraction of hot white dwarfs with a low mass companion is derived. 
In Sect.\ \ref{sec:QSO} the matches of the UV-excess sources in the \WISE data are presented, and additionally a 
list of candidate QSOs with $|b|$$<$5\degr is selected. Finally in Sect.\ \ref{sec:discussion} we summarise and 
discuss the conclusions.\\

\begin{figure*}
\centerline{\epsfig{file=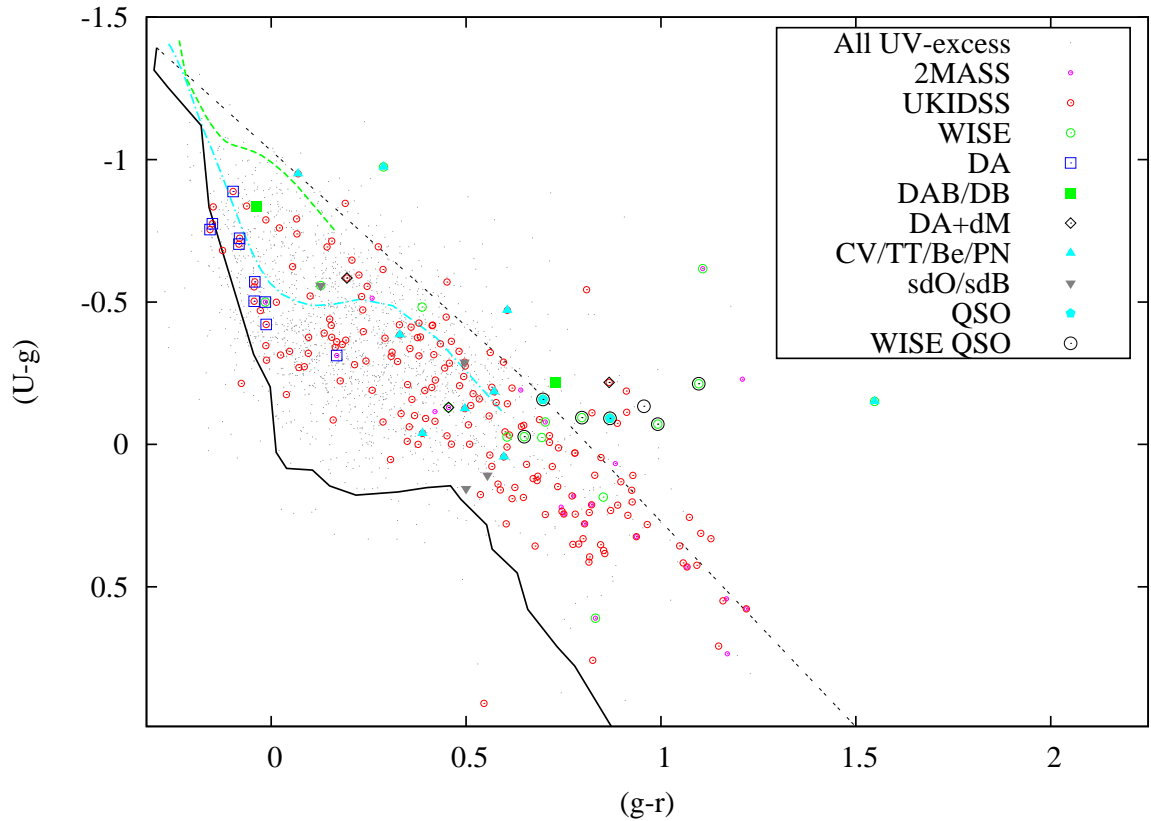,width=16cm,angle=0,clip=}}
\caption{Colour-colour diagram with the UV-excess matches in UKIDSS, 2MASS and WISE.
UV-excess sources spectroscopically classified in V12b are overplotted with different symbols. 
The lines are the simulated colours of unreddened main-sequence stars (solid black) and the O5V-reddening line (dashed black) of V12a. 
The cyan and green dashed lines are respectively the simulated colours of unreddened Koester DA and DB white dwarfs.
The grey dots are the sources from the complete UV-excess catalogue of V12a.
\label{fig:UVEXccdIR}}
\end{figure*}

\begin{figure*}
\centerline{\epsfig{file=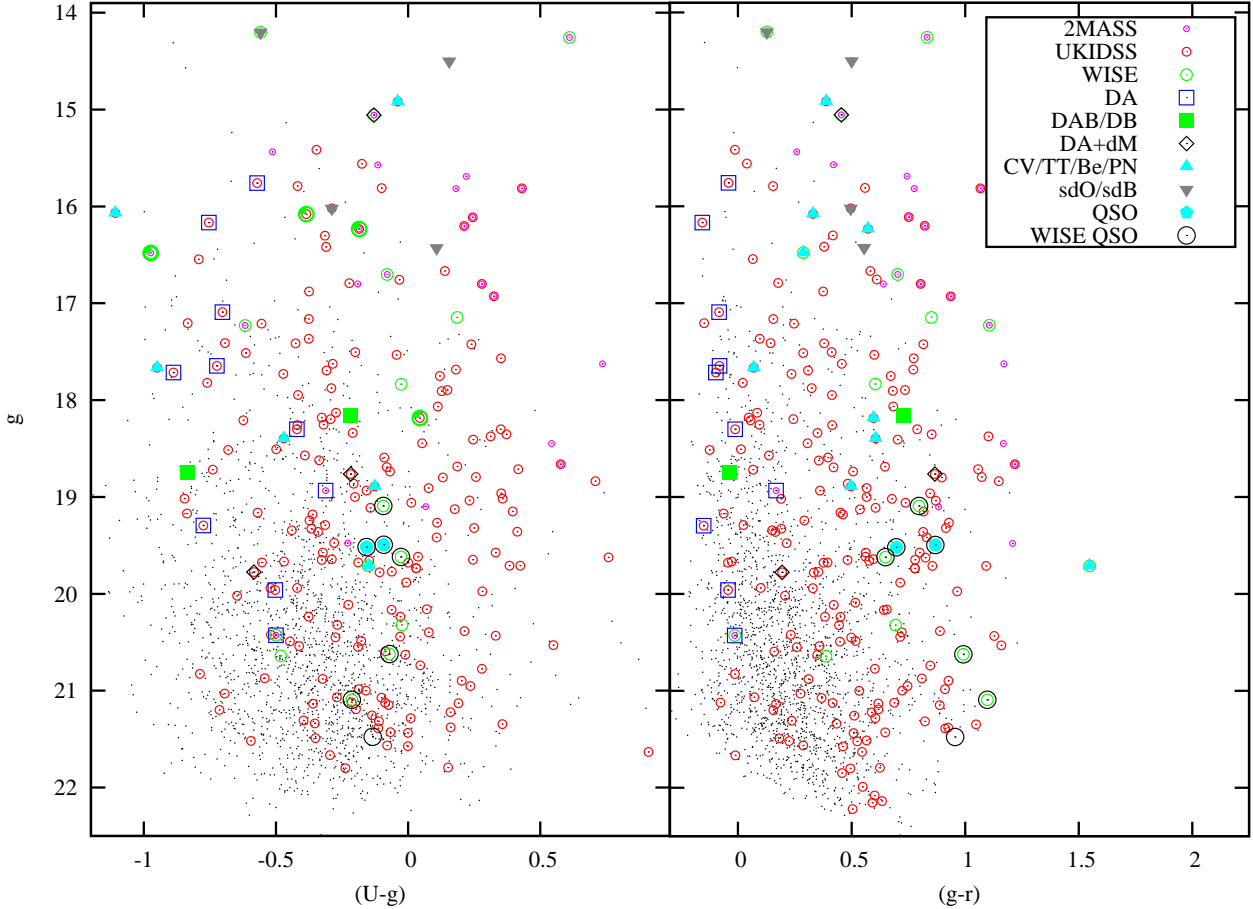,width=22cm,angle=0,clip=}}
\caption{Colour-magnitude diagrams with the UV-excess matches in UKIDSS, 2MASS and WISE.
Spectroscopically identified UV-excess sources of V12b are overplotted with different symbols.
The grey dots are the sources from the complete UV-excess catalogue of V12a.
\label{fig:UVEXcmdIR}}
\end{figure*}

\begin{figure*}
\centerline{\epsfig{file=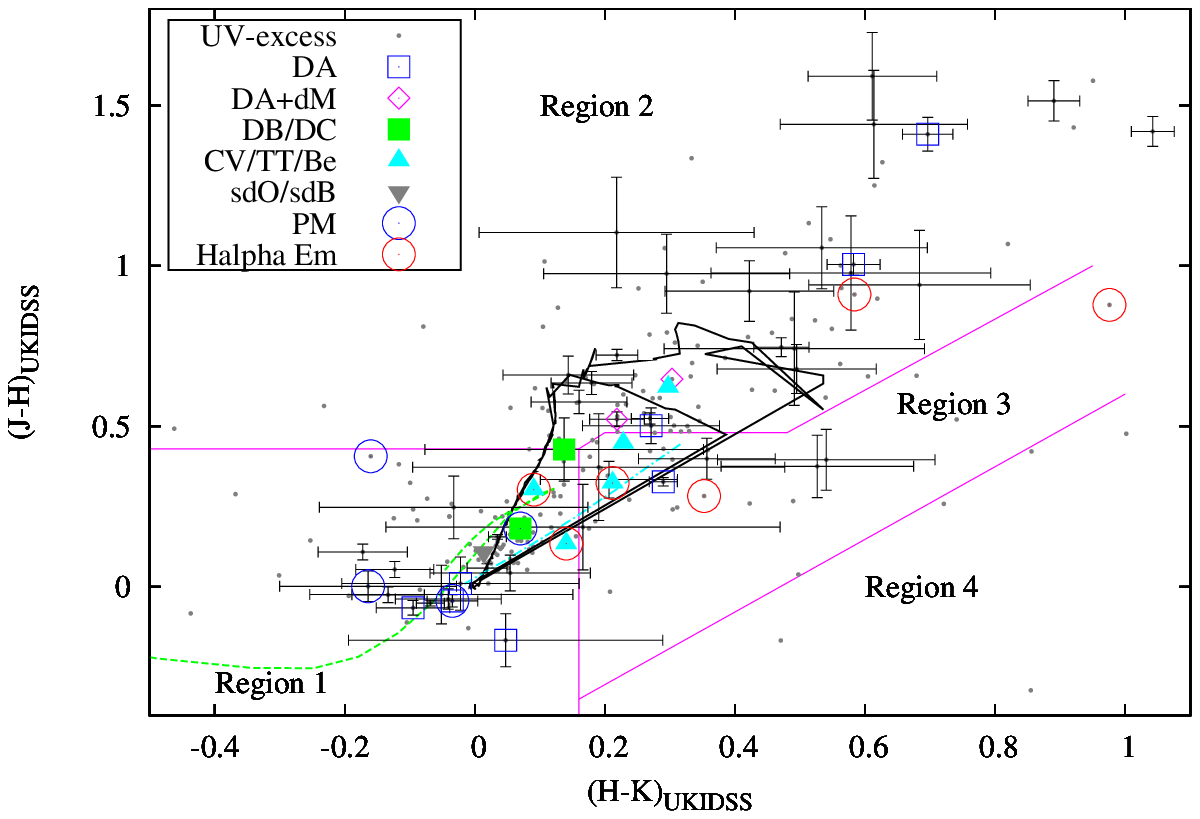,width=16cm,angle=0,clip=}}
\caption{The $(J-H)$ vs. $(H-K)$ colour-colour diagram with the UKIDSS-GPS matches. The tracks are the unreddened 
UKIDSS colours of main-sequence stars (black), DA white dwarfs (green) and DB white dwarfs (cyan).
UV-excess sources spectroscopically classified in V12b are overplotted with different symbols,
UV-excess candidate white dwarfs with $(g-r)<$0.2 are plotted with error bars, 
other UV-excess sources are plotted with dots. 
The four regions are indicated that contain different candidates: 
1) single white dwarfs, 2) white dwarfs with an M-dwarf companion,
3) candidate white dwarfs with a later type (brown dwarf) companion, and 
4) white dwarf with circumstellar material or a debris disk.
\label{fig:UKIDSSccd}}
\end{figure*}

\section{Cross-matching with IR surveys: UKIDSS, 2MASS and WISE}
\label{sec:crossmatching}
\begin{figure*}
\centerline{\epsfig{file=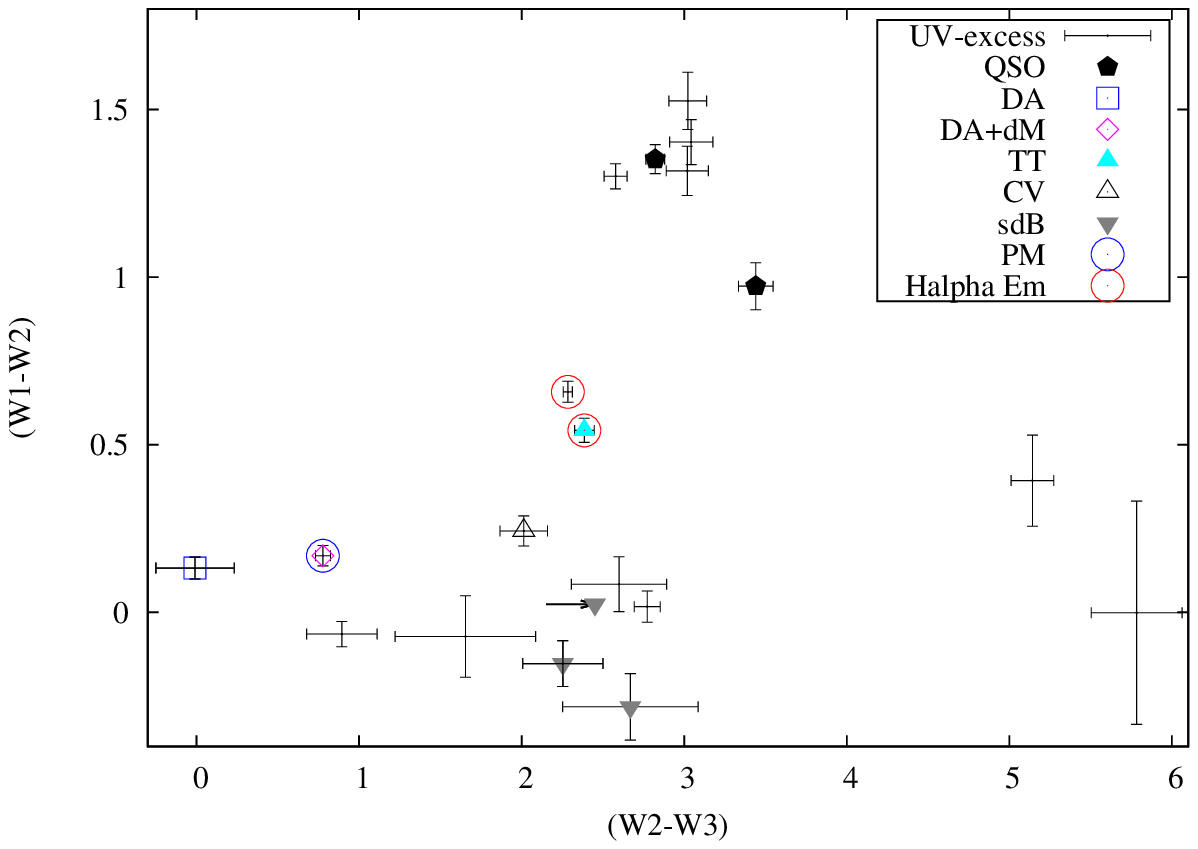,width=16cm,angle=0,clip=}}
\caption{WISE matches within 1 arcsec plotted in the WISE colour-colour diagram. 
Sources which are classified from their spectra in V12b are overplotted with different symbols, 
all UV-excess matches are plotted with error bars. The 2 sources at $(W2-W3)$=0 and $(W2-W3)$=5.8 are 
UV-excess candidate white dwarfs with $(g-r)<$0.2, 1 classified as DA white dwarf in V12b and 
1 unclassified source with a strong $W3$ excess and $(W3-W4)$=2.1.
\label{fig:WISEccd}}
\end{figure*}

The UV-excess catalogue of V12a (2\,170 sources) is cross-matched with different surveys that image (parts of) 
the Galactic Plane at red/infrared (IR) wavelengths.
An overview of the cross-matching is given in Table\ \ref{tab:crossmatching}.
Note that the coverage of the Galactic Plane and the overlap with the \UVEX fields of V12a is not complete for 
all surveys (see Sect.\ \ref{sec:discussion}).
The results of the cross-matching are shown in the colour-colour 
diagrams of Figs.\ \ref{fig:UVEXccdIR} to\ \ref{fig:WISEccd}, where the 
spectroscopically classified sources of V12b are labelled.
The \UKIDSS and 2\MASS colour-magnitude diagrams are shown in the Appendix.
As expected, in particular the redder and brighter \UVEX 
UV-excess sources have a larger fraction of IR matches, as can be seen in the \UVEX colour-colour and colour-magnitude diagrams
of Figs.\ \ref{fig:UVEXccdIR} to\ \ref{fig:UVEXcmdIR}.
The different IR surveys and the number of matches with the full UV-excess catalogue of V12a are described below.\\

\begin{table}
\caption[]{Summary of the cross-matching: No of matches with the full UV-excess catalogue
and No of matches in the Deacon PM/Witham H$\alpha$ catalogues (Deacon et al., 2009, Witham et al., 2008). \label{tab:crossmatching} }
\centering
{\small
\begin{tabular}{ | l | c | c | }
    \hline
Catalogue:                 &  Full UV-excess      &  PM/H$\alpha$ \\ \hline	   
UKIDSS-GPS                 &  227   	          &  4/6       \\
2MASS                      &  60  	          &  2/5       \\
WISE                       &  19    	          &  1/2       \\ 
IPHAS-IDR                  &  1\,203              &  26/15     \\    
SDSS DR8                   &  378   	          &  6/3       \\
    \hline
\end{tabular} \\ 
}
\end{table}

\begin{figure*}
\centerline{\epsfig{file=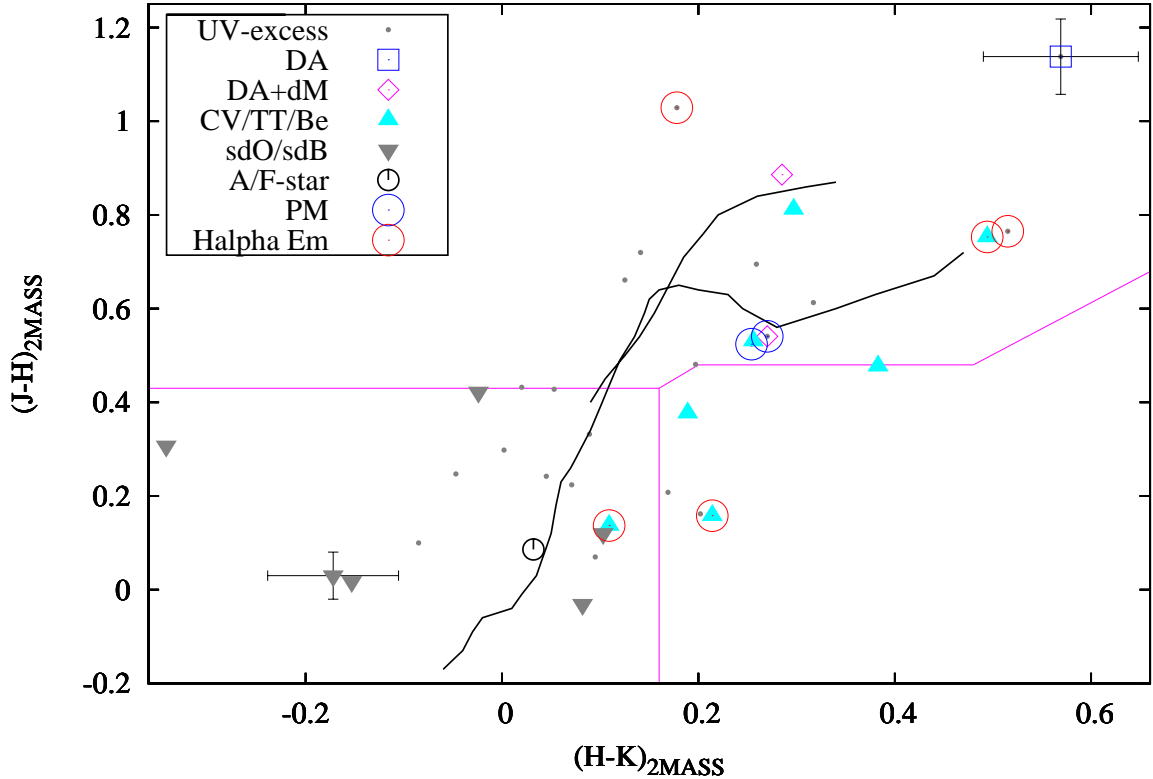,width=16cm,angle=0,clip=}}
\caption{$(J-H)$ vs. $(H-K)$ colour-colour diagram with the UV-excess matches in 2MASS.
The black lines are the simulated colours of main-sequence stars and giants with $E(B-V)$=0. 
Classified sources are labelled with different symbols, UV-excess candidate white dwarfs are plotted with error bars, 
other UV-excess sources are plotted with dots.
There is one more match at $(H-K)$=1.15, $(J-H)$=2.6, classified as DA white dwarf in V12b, not visible in this figure.
\label{fig:2MASSccd}}
\end{figure*}

\begin{itemize}

\item The UKIRT InfraRed Deep Sky Survey (\UKIDSS, Lawrence et al., 2007) is a near-infrared survey imaging the northern sky
in the $J$, $H$ and $K$ (1.2, 1.6 and 2.2 micron) filters using the Wide Field Camera (WFCAM) mounted on 
the 3.8m United Kingdom Infra-red Telescope (UKIRT) on Hawaii.
The \UKIDSS Galactic Plane Survey (\UKIDSS GPS, Lucas et al., 2008, Lawrence et al., 2012) images the northern Galactic Plane in the same Galactic latitude
range as \UVEX and \IPHAS.
There is a match for a total of 227 UV-excess sources in all three \UKIDSS filters 
within a radius of 1 arcsec (10$\%$ of the complete UV-excess catalogue).
Note that the overlap of the \UKIDSS GPS DR1 with the \UVEX fields of V12a is not complete.
The UV-excess 3-filter matches in \UKIDSS with $K$$>$11
are plotted in the colour-colour diagram of Fig.\ \ref{fig:UKIDSSccd} 
on top of the simulated unreddened main-sequence colours (Hewett et al., 2006).
Of these \UKIDSS matches there are 18 sources spectroscopically classified in V12b, 4 matches are in the IPHAS-POSSI 
proper motion catalogue (Deacon et al., 2009) and 6 matches are in the H$\alpha$ emitter catalogue (Witham et al., 2008).
\\

\item The Two-Micron All-Sky Survey (2\MASS, Skrutskie et al., 2006, Cutri et al., 2003) imaged the entire sky in
the three near-infrared filter bands $J$, $H$ and $K$ (1.2, 1.7 and 2.2 micron)
with a limiting magnitude of $J$=17.1, $H$=16.4 and $K$=15.3, using 2 automated 1.3m telescopes, 
one at Mt. Hopkins, Arizona, and one at CTIO, Chile.
The overlap of the 2\MASS All-Sky Catalog of Point Sources (Cutri et al., 2003) data and 
the \UVEX fields of V12a is 100$\%$.
In 2\MASS there is a match for 60 sources of the complete UV-excess catalogue 
in all three 2\MASS filter bands within a radius of 1 arcsec (3$\%$).
Eighteen of these matches were spectroscopically classified in V12b, 2 matches are the IPHAS-POSS catalogue (1 classified as DA+dM)
and 5 matches are in the H$\alpha$ emitter catalogue (3 classified as T Tauri, Be star and Cataclysmic Variable).
The UV-excess matches in 2\MASS are plotted in the colour-colour diagram of Fig.\ \ref{fig:2MASSccd}.
\\

\item The Wide-field Infrared Survey Explorer (\WISE, Wright et al., 2010) mapped the sky at four mid-infrared bands 
at 3.4, 4.6, 12, and 22 micron ($W1$, $W2$, $W3$ and $W4$). The overlap of the \WISE All-Sky Data Release (Cutri et al., 2012) 
and the \UVEX fields of V12a is 100$\%$. The \WISE data are used to select and classify UV-excess sources with a mid-IR excess, 
and a list of UV-excess candidate QSOs is selected in Sect.\ \ref{sec:QSO}.
There is a match for 20 UV-excess sources in \WISE
within a radius of 1 arcsec in at least the first three filters (W1, W2, W3).
Thirteen of them have a match in all four \WISE filters. 
The 3-filter matches are shown in the colour-colour diagram of Fig.\ \ref{fig:WISEccd}.
\\

\item The INT/WFC Photometric H$\alpha$ Survey of the Northern Galactic Plane (\IPHAS, Drew et al., 2005) has imaged the
same survey area as \UVEX with the same telescope and camera set-up using the $r$, $i$ and $H\alpha$ filters. 
There is a match for 1\,203 of our 2\,170 UV-excess sources in the \IPHAS initial data release (IDR, Gonz\'{a}lez-Solares et al., 2008)
within a radius of 1.0 arcsec (55$\%$). 
Note that the overlap of the \IPHAS IDR with the \UVEX fields of V12a is not complete, but $\sim$90$\%$.
The result of the cross-match between the UV-excess catalogue and \IPHAS IDR was already shown in Figs.\,10 and 11 of V12a and Fig.\,3 of V12b.
If available, the \IPHAS data is used to distinguish UV-excess white dwarfs from UV-excess sources showing H$\alpha$ emission, and
the $i$-band photometry is used in the spectral fitting in Sect.\ \ref{sec:fitting}.
In the colour-colour diagrams of Figs.\ \ref{fig:UKIDSSccd} to\ \ref{fig:WISEccd} the sources
that are in the \IPHAS H$\alpha$ emitter catalogue (Witham et al., 2008) or IPHAS-POSS proper motion (PM) catalogue 
(Deacon et al., 2009) are circled red and blue respectively. The Witham H$\alpha$ emitter catalogue covers the magnitude range 
13$<$$r$$<$19.5 and the Deacon IPHAS-POSS PM catalogue covers the magnitude range 13.5$<$$r$$<$19.
Matches in the H$\alpha$ emission line star catalogue are expected to be Galactic sources, except for QSOs with redshift 0.5 or 1.3,
which have emission lines exactly in the bandpass of the \IPHAS H$\alpha$ filter (Scaringi et al., 2013).
\\

\item Additionally, the Sloan Digital Sky Survey (\SDSS, York et al., 2000) Photometric Catalog DR 8 (Adelman-McCarthy et al., 2011) 
overlaps with some \UVEX Galactic Plane fields. \SDSS images the sky in the filters $u$, $g$, $r$, $i$, $z$ down 
to $\sim 22^{nd}$ magnitude, using the 2.5m wide-angle optical telescope at Apache Point Observatory, New Mexico, US.
For the full UV-excess catalogue 378 sources have a match in \SDSS 
within a radius of 1 arcsec. Note that the overlap between \SDSS and the \UVEX fields of V12a is not complete.
There are 6 \SDSS UV-excess matches in the IPHAS-POSS PM catalogue and 3 \SDSS UV-excess matches are in the \IPHAS H$\alpha$ emitter catalogue.
For the UV-excess sources that have an \SDSS match, the additional $i$-band and $z$-band photometry is used for the 
spectral fitting (Fig.\,1, Rebassa-Mansergas et al., 2012) in Sect.\ \ref{sec:fitting}.
\\

\end{itemize}

\section{Candidate white dwarfs with an infrared counterpart}
\label{sec:WDIR}
The UV-excess catalogue consists of a mix of different populations. From the spectroscopic follow-up of V12b, 
52$\%$ of the UV-excess sources are single DA white dwarfs, 14$\%$ are white dwarfs of other types (DAB/DB/DC/DZ/DAe), 
4$\%$ are DA+dM white dwarfs, 11$\%$ are sdB/sdO stars, 9$\%$ are H$\alpha$ emission line objects, 8$\%$ 
are BHB/MS stars and 2$\%$ are QSO.
In Sects.\ \ref{sec:UKIDSS2MASS} and\ \ref{sec:fitting} we focus in particular on hot white dwarfs with an IR-match.
For that reason a sub-sample of 964 UV-excess candidate white dwarfs
with $(g-r)$$<$0.2 is selected, which corresponds with the simulated unreddened colours of DA white dwarfs hotter than $T_{\rm eff}$$>$9\,000K.
This sub-sample does not contain DA white dwarfs cooler than $T_{\rm eff}$$<$9\,000K or strongly reddened white dwarfs.
From spectroscopic follow-up (Fig.\,1, V12b) it is known that 97$\%$ of the DA white dwarfs 
identified in \UVEX are in this colour range, but there will be $\sim$25$\%$ 
white dwarfs of other types and sdO/sdB stars which also may have infrared counterparts. 
Only the IR-matches of the sources from this UV-excess candidate white dwarf sub-sample are plotted with error bars in the colour-colour
diagrams of Figs.\ \ref{fig:UKIDSSccd} to\ \ref{fig:WISEccd}.\\

\subsection{Classification of white dwarfs in UKIDSS and 2MASS}
\label{sec:UKIDSS2MASS}
In the hot sub-sample there are 46 UV-excess candidate white dwarfs with a \UKIDSS match, and 3 with a 2\MASS match. These matches are 
plotted with error bars in the colour-colour diagrams of Figs.\ \ref{fig:UKIDSSccd} and\ \ref{fig:2MASSccd}.
To separate single white dwarfs from white dwarfs with a companion or white dwarfs with a debris disk the $(J-H)$ vs. $(H-K)$ colour-colour diagram 
is divided in four separate regions following Wachter et al. (2003) and Steele et al. (2011). 
The different regions in Fig.\ \ref{fig:UKIDSSccd} contain different candidates:\\

\begin{itemize}

\item Region 1:\,
There are 17 UV-excess sources that are single candidate white dwarfs, of which 4 were already classified as 
hydrogen atmosphere (DA) white dwarfs and 1 classified as sdB star in V12b.

\item Region 2:\,
Sources in region 2 are candidates for white dwarfs with an M-dwarf companion.
There are 22 UV-excess candidates, of which 11 have $g$$<$20.
Five of these sources are classified as single DA white dwarfs in V12b. 
However, the available spectra cannot exclude the presence of a late-type companion.

\item Region 3:\,
Sources in region 3 are candidates for white dwarfs with a later type (brown dwarf) companion.
There are 7 UV-excess candidate white dwarfs, one of them is classified as a DA white dwarf in V12b (UVEXJ202659.21+411644.1). 
The available spectrum cannot exclude the presence of a very late-type companion.

\item Region 4:\,
Sources in region 4 show a K-band excess, possibly due to circumstellar material or a disk.
There are 5 UV-excess sources, but none of them are candidate hot white dwarfs since they have $(g-r)$$>$0.2 in \UVEX. 

\end{itemize}

\subsection{Determination of Spectral Types}
\label{sec:fitting}
For the UV-excess candidate white dwarfs with an IR-excess match in \UKIDSS and/or 2\MASS the Spectral Energy Distributions (SEDs) are fitted
in order to determine the spectral types. 
Grids of DA+dM model spectra, in the range 0$<$$E(B-V)$$<$1.0 at $E(B-V)$=0.1 intervals, using the reddening laws of
Cardelli, Clayton \& Mathis (1989), are fitted to the optical and infrared
photometry in order to determine white dwarf temperatures. 
For the fitting the white dwarf atmosphere models and M-dwarf models are both placed at the same distance.
The spectral fluxes of Beuermann, 2006 are used to calibrate the absolue fluxes of the M-dwarfs.
The grid of DA+dM model spectra is constructed from white dwarf atmosphere model spectra of Koester 
(2001) with $log\,g$=8.0 in the range 6\,000$<$$T_{\rm eff}$$<$80\,000K, and template spectra of main-sequence stars from the library of Pickles (1998) 
with spectral type M0V to M6V.
First, the \UVEX photometry is used to determine the temperature of the white dwarf.
For the DA+dM SED fitting the photometry of \UVEX, \UKIDSS and 2\MASS, and 
additionally if available, the \IPHAS and/or \SDSS photometry is used for 40 candidate white dwarfs.
The \WISE photometry is not used since the wavelength range of the DA+dM models only covers 3\,000-25\,000 \AA, 
but consistency with the \WISE photometry was checked, see e.g. Fig.\ \ref{fig:SED7}.\\

The DA+dM models do not give a good fitting result for all candidate white dwarfs with an IR-excess match.
Some UV-excess sources with an IR-excess match can be fully explained by a reddened sdB or sdO spectrum without any companion, or
the IR-excess can be explained by an sdB/sdO stars with an F-, G- or K-type main-sequence companion.
SdB/sdO stars with later companions can not be identified with the current photometry because the sdB/sdO dominates the SED out to the $K$-band.
Grids of TheoSSA (Ringat, 2012) sdB/sdO models with $log\,g$=5.5 in the range 20\,000$<$$T_{\rm eff}$$<$50\,000K and 0$<$$E(B-V)$$<$5.0 
at $T_{\rm eff}$=1\,000K and $E(B-V)$=0.1 intervals are fitted to the optical and infrared
photometry to determine $T_{\rm eff}$ and the reddening of the sdB/sdO stars. For the sdO/sdB stars with a possible main-sequence companion, a grid
is constructed from the reddened sdB/sdO models and the template spectra of main-sequence stars from the library of Pickles (1998) 
with spectral types A5V to M5V. Also here the \WISE photometry is not used, but consistency was checked, since the sdO/sdB+MS models 
cover the wavelength range 3\,000-25\,000 \AA.\\

Candidate white dwarfs in region 3 of the IR colour-colour diagrams have companions with spectral types later than M6. 
The spectral types of these companions are determined using the $(J-K)$ colours resulting after subtracting the white dwarf flux
as explained in Reid et al., 2001 and Leggett et al., 2002. The \UKIDSS colours are converted to AB colours taking 
into account the correction $(J-K)$=0.962 of Hewett et al. (2006).\\

In the original UV-excess catalogue there might be a possible systematic shift in the \UVEX $U$-band data, which would influence
the result of Spectral Energy Distributions fitting in this paper. 
For that reason recalibrated \UVEX data, as explained in Greiss et
al. (2012), is used. The shift in the original \UVEX data does not influence the content of the UV-excess catalogue 
because the selection in V12a was done relative to the reddened main-sequence population. 
The magnitudes and colours of the UV-excess sources might still show a small scatter,
similar to the early \IPHAS data (Drew et al., 2005), since a global photometric calibration is not applied to the \UVEX data.\\

To convert the magnitudes into fluxes ($F$) we use: 
$F$ = 10$^{-0.4\times(mag_{(AB)}+48.6)}$ for \EGAPS and \SDSS.
For \EGAPS photometry the AB offsets $U$=0.927, $g$=-0.103, $r$=0.164 and $i$=0.413 need to be added to convert to AB magnitudes
(Gonz\'{a}lez-Solares et al., 2008; Blanton \& Roweis, 2007 and Hewett et al., 2006).
To convert \UKIDSS photometry to AB magnitudes the correction $J$=0.938, $H$=1.379 and $K$=1.900 need to be taken into 
account (Hewett et al., 2006).
For 2\MASS the flux is derived using $F$ = $F_{\nu} - 0\,mag$ $\times$ 10$^{-0.4\times(mag_{(Vega)})}$, 
where $F_{\nu} - 0\,mag$ is 1594, 1024 and 666,7 (Jy) for $J$, $H$ and $K$ 
respectively (Cutri et al., 2003).
The \WISE photometry is converted into fluxes using $F$ = $F_{\nu 0}$ $\times$ 10$^{-0.4\times mag_{(Vega)}}$,
where $F_{\nu 0}$  is the the ``Zero Magnitude Flux Density'' for the \WISE filter bands: 
$W1$=309.54, $W2$=171.787, $W3$=12.82, $W4$=9.26 (Jy) (Wright et al. 2010).\\

\subsection{Fitting results}
\label{sec:fittingresults}
\begin{figure}
\centerline{\epsfig{file=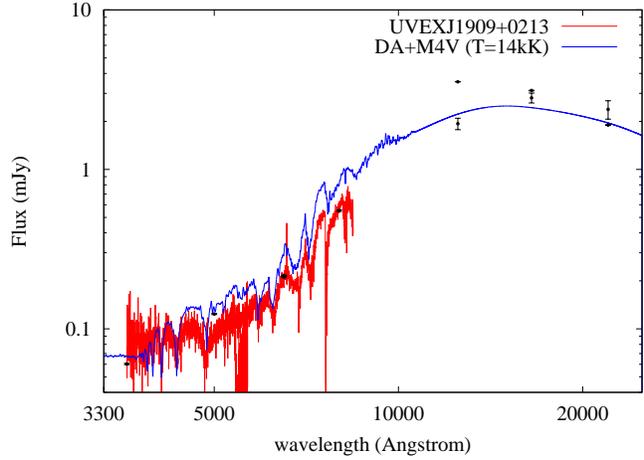,width=9cm,angle=0,clip=}}
\caption{The SED of UVEXJ190912.34+021342.8, classified as DA+dM in V12b, with overplotted the best DA+dM model (blue), 
a $T_{\rm eff}$=14kK white dwarf plus M4V companion.
Plotted here are the UVEX, IPHAS, 2MASS and UKIDSS photometry with error bars and the WHT spectrum of V12b (red).
\label{fig:SED6}}
\end{figure}

\begin{figure}
\centerline{\epsfig{file=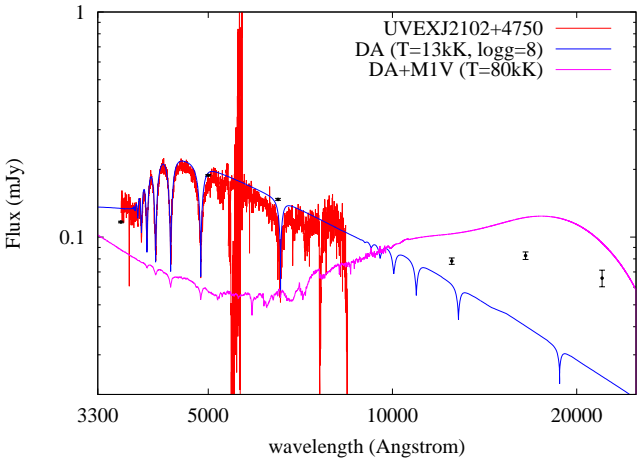,width=9cm,angle=0,clip=}}
\caption{The SED of UVEXJ210248.44+475058.9, classified as DA white dwarf with $T_{\rm eff}$=13.3kK and $log(g)$=8.1 in V12b, 
shows a clear excess in the infrared which can be explained by a companion. 
The best-fit DA+dM model is a $T_{\rm eff}$=80kK white dwarf plus M1V companion (magenta), which is clearly too hot for the white dwarf.
From the $(J-K)$ colour the spectral type of the companion is L5.
Plotted here are the UVEX and UKIDSS photometry with error bars, the WHT spectrum of V12b (red) and a
Koester DA white dwarf atmosphere model with $T_{\rm eff}$=13kK and $log(g)$=8.0 (blue).
\label{fig:SED1}}
\end{figure}

To test our photometric fitting routine first the method is applied to the UV-excess candidate white dwarfs that were spectroscopically
classified in V12b. Note that the aim of the fitting is not to derive accurate temperatures, spectral types and reddening, but to classify
the sources, and to confirm (or to rule out) the presence of a companion or disk.
Fitting results for 7 UV-excess sources, spectroscopically classified in V12b, with an IR-excess or just an IR match, 
are shown in Figs.\ \ref{fig:SED6} to\ \ref{fig:SED1}.
For some of these sources there is also a match in \WISE (see Sect.\ \ref{sec:WISEwhitedwarfs}).\\

\begin{figure}
\centerline{\epsfig{file=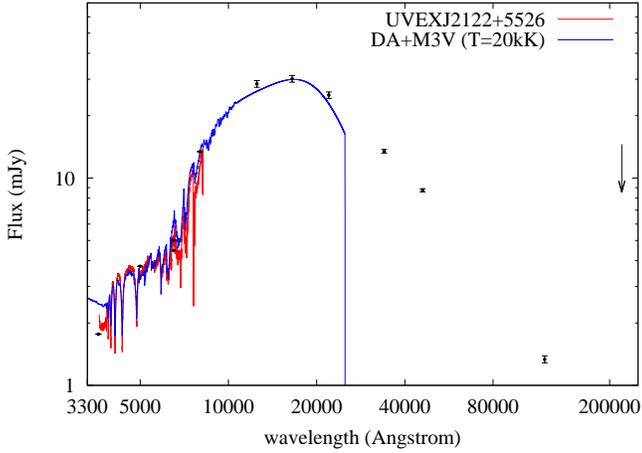,width=9cm,angle=0,clip=}}
\caption{The SED of UVEXJ212257.82+552609.0, classified as DA+dM in V12b, can indeed be explained by DA+dM model (blue): a
$T_{\rm eff}$=20kK white dwarf plus M3V companion, except for the W4 WISE photometry which is spurious.
Plotted here are the UVEX, IPHAS, 2MASS and WISE photometry with error bars and the WHT spectrum of V12b (red).
\label{fig:SED7}}
\end{figure}

\begin{figure}
\centerline{\epsfig{file=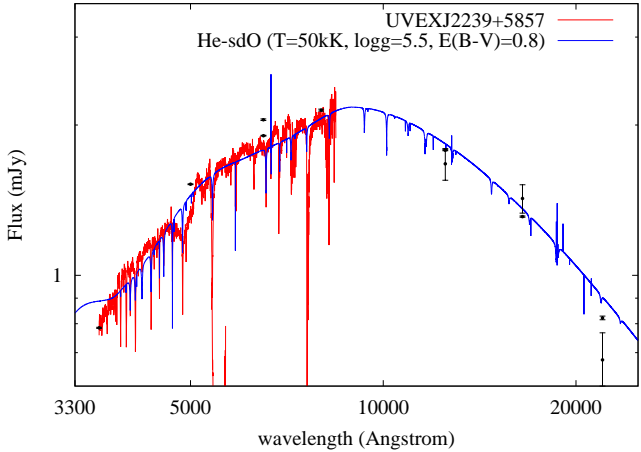,width=9cm,angle=0,clip=}}
\caption{The photometry of UVEXJ223941.98+585729.1, classified as He-sdO in V12b, can be fully explained by a reddened He-sdO spectrum with 
$T_{\rm eff}$=50kK, $log(g)$=5.5 and $E(B-V)$=0.8 (blue). Plotted here are the UVEX, IPHAS, UKIDSS and 2MASS photometry with error bars and
the WHT spectrum of V12b (red).
\label{fig:SED4}}
\end{figure}

\begin{figure}
\centerline{\epsfig{file=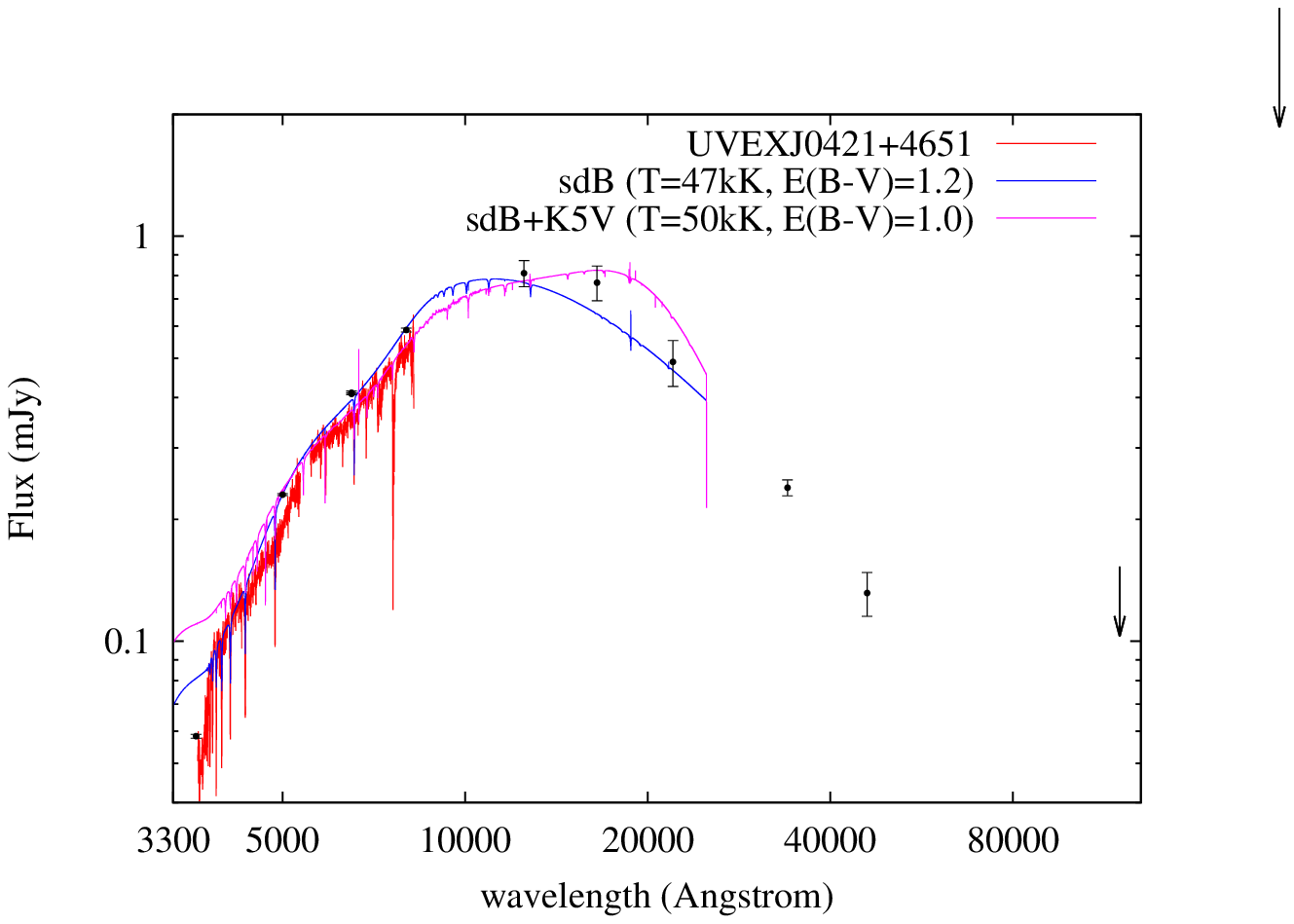,width=9cm,angle=0,clip=}}
\caption{The photometry of UVEXJ042125.70+465115.4, classified as sdB+F in V12b, 
overplotted with a single sdB spectrum with $T_{\rm eff}$=47kK, $E(B-V)$=1.2 (blue), 
and the best-fit sdB+MS model: a sdB+K5V spectrum with $T_{\rm eff}$=50kK, $E(B-V)$=1.0 (magenta).
Plotted here are the UVEX, IPHAS, 2MASS and WISE photometry with error bars and the WHT spectrum of V12b (red).
\label{fig:SED5}}
\end{figure}

\begin{figure}
\centerline{\epsfig{file=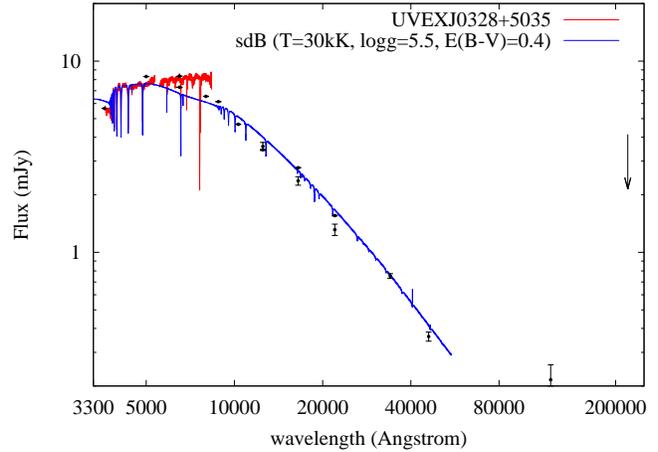,width=9cm,angle=0,clip=}}
\caption{The photometry of UVEXJ032855.25+503529.8 can be fully explained by a reddened sdB model 
spectrum (blue) with $T_{\rm eff}$=30kK, $log(g)$=5.5 and $E(B-V)$=0.4. 
Plotted here are the UVEX, IPHAS, UKIDSS, 2MASS and WISE photometry with error bars and the WHT spectrum of V12b (red).
The WHT spectrum might deviate at the red end of the spectrum ($\lambda>$7\,500 \AA ) due to the flux calibration.
\label{fig:SED3}}
\end{figure}

\begin{figure}
\centerline{\epsfig{file=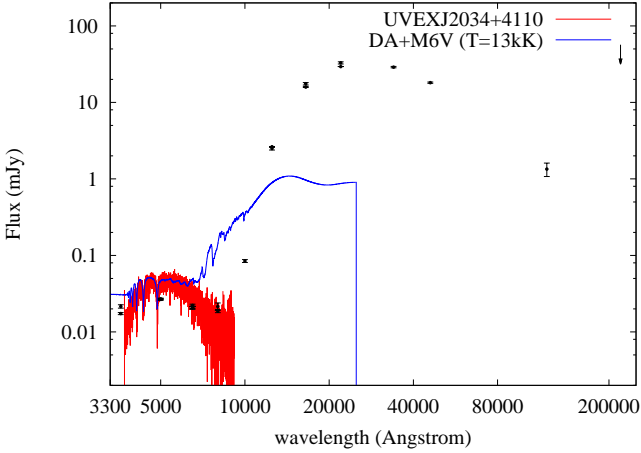,width=9cm,angle=0,clip=}}
\caption{The SED of UVEXJ203411.72+411020.3, classified as single DA white dwarf in V12b, can not be explained by one of our DA 
white dwarf plus a late type companion models. The best-fit DA+dM model is a $T_{\rm eff}$=13kK white dwarf plus M6V companion (blue). 
The strong infrared-excess in 2MASS, UKIDSS and WISE can be explained by a later companion with spectral type L8. 
However, the contribution of this L8 dwarf is significantly more luminous than expected, so the WD and BD are expected to be at 
different distances and not physically associated.
Plotted here are the UVEX, IPHAS, SDSS, 2MASS, UKIDSS and WISE photometry with error bars 
and the Hectospec spectrum of V12b (red).
\label{fig:SED2}}
\end{figure}

\begin{itemize}

\item The object UVEXJ1909+0213, classified as DA+dM objects in V12b, is shown in Fig.\ \ref{fig:SED6}.
This source has $(g-r)$=0.87, so it is not in the UV-excess candidate white dwarf sub-sample. 
The best-fit model consists of a white dwarf with $T_{\rm eff}$=14kK with an M4V companion.
This source is in region 2 of the 2\MASS colour-colour diagram of Fig.\ \ref{fig:2MASSccd} and in region 2 of 
the \UKIDSS colour-colour diagram of Fig.\ \ref{fig:UKIDSSccd}. Note that only the photometry was used for the fitting.

\item The object UVEXJ2122+5526, also classified as DA+dM objects in V12b, is shown in Fig.\ \ref{fig:SED7}.
This source has $(g-r)$=0.47, so it is also not in the UV-excess candidate white dwarf sub-sample. 
The model that fits best consists of a white dwarf with $T_{\rm eff}$=20kK with an M3V companion, 
which is consistent up to the $W3$ \WISE photometry.
This source is in region 2 of the 2\MASS colour-colour diagram of Fig.\ \ref{fig:2MASSccd} and
is plotted in the \WISE colour-colour diagram of Fig.\ \ref{fig:WISEccd}.

\item The SED of UVEXJ2239+5857, classified as He-sdO in V12b, shows a decreasing flux in the IR which can be explained with a reddened sdB/sdO spectrum 
with $T_{\rm eff}$=50kK, $log\,g$=5.5 and $E(B-V)$=0.8, which is the model that fits best for this object, see Fig.\ \ref{fig:SED4}.

\item The SED of UVEXJ0421+4651, classified as sdB+F in V12b, can be explained by a single reddened sdB/sdO spectrum, but the combination of a
sdB and a K5V spectrum gives a slightly better fit with $T_{\rm eff}$=50kK, $E(B-V)$=1.0, see Fig.\ \ref{fig:SED5}.
This source was classified as sdB+F in V12b due to the CaII lines present in the optical spectrum.

\item The photometry of UVEXJ0328+5035, classified as a sdB star in V12b, can be fully explained by a reddened sdB spectrum with 
$T_{\rm eff}$=30kK, $log\,g$=5.5 and $E(B-V)$=0.4, see Fig.\ \ref{fig:SED3}. However, this object is known to be a sdB+dM binary from its radial velocity
(Kupfer et al., in prep.). No sign of the companion is seen in the SED.

\item The SED fitting method finds a DA+dM as best solution for source UVEXJ2034+4110 (Fig.\ \ref{fig:SED2}), classified as a DA white dwarf in V12b. 
The best-fit DA+dM model is a $T_{\rm eff}$=13kK DA white dwarf plus an M6V companion. 
The strong infrared-excess is due to a low-mass companion with spectral type L8 as determined from the $(J-K)$ colour.
However, the contribution of an low-mass companion with spectral type L8 would be less luminous than in Fig.\ \ref{fig:SED2},
so the white dwarf is probably at a larger distance compared to the brown dwarf.
Additionally, the blue/optical images have a match within 0.1 arcsec, while there is an offset for the infrared images of 0.6 arcsec.
The white dwarf and L8 brown dwarf are expected to be at different distances and not physically associated.

\item The source UVEXJ2102+4750 in Fig.\ \ref{fig:SED1}, classified as a DA white dwarf with $T_{\rm eff}$=13.3kK and $log(g)$=8.1 in V12b, 
shows an IR-excess. None of the DA+dM models fit the photometry well, 
making it candidate for having a companion with spectral type L5 as determined from the $(J-K)$ colour.
This source was classified as a DA white dwarf in V12b since there is no sign of the companion in the optical spectrum, 
which is consistent with the IR-excess which increases for wavelengths larger than $\lambda$$>$8\,000 \AA.

\end{itemize}

The results of the fitting for all UV-excess candidate white dwarfs with an IR-excess match in \UKIDSS and/or 2\MASS are shown 
Table\ \ref{tab:IRexcess2} of the Appendix\ \ref{app:appendix}. Note that $T_{\rm eff}$ of the white dwarf and the spectral 
types of the companions are rough estimates, since only photometry is used for the fitting.
From the positions in the $(J-H)$ vs. $(H-K)$ colour-colour diagram, and from the SED 
fitting, 24 UV-excess candidate white dwarfs are classified as white dwarf with an M-dwarf companion, 
7 sources are candidate white dwarfs probably with a brown dwarf type companion (later than M-type),
19 UV-excess candidate white dwarfs are single white dwarfs or single sdO/sdB stars without a companion,
and no UV-excess white dwarfs are clear debris disk candidates.\\

\subsection{UV-excess sources in the Wide-field Infrared Survey data}
\label{sec:WISEwhitedwarfs}
\begin{figure}
\centerline{\epsfig{file=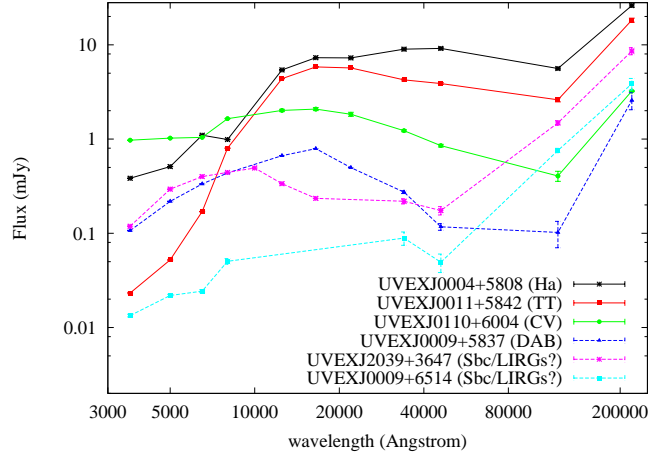,width=9cm,angle=0,clip=}}
\caption{The SEDs of the UV-excess sources classified in V12b with a match in WISE and
the 2 unclassified sources at $(W2-W3)\sim$5.5 which are candidate star-burst Sbc/LIRGs (dashed lines).
\label{fig:SEDclassified}}
\end{figure}

The release of the all-sky \WISE catalog (Cutri et al., 2012) contains all sky data in four mid-infrared bands centered 
at 3.4, 4.6, 12 and 22 micron. 
There are 9 classified UV-excess sources with a \WISE match, 1 classified as DA white dwarf, 1 classified as Cataclysmic Variable, 
1 classified as DA+dM, 1 classified as T Tauri star, 3 classified as sdB stars, and 2 classified as QSOs in V12b.
The QSOs and 4 new QSO candidates are discussed in Sect.\ \ref{sec:QSO}.
The SEDs of the sources classified as DA+dM (UVEXJ2122+5526), DA (UVEXJ2034+4110; ccd-flag ``H'' for $W3$), 
sdB+F (UVEXJ0421+4651; $W3$ and $W4$ are upper limits), 
and sdB (UVEXJ0328+5035) in V12b, were already shown in Figs.\ \ref{fig:SED7} to\ \ref{fig:SED2}.
The SEDs of objects of other types, classified in V12b, with a match in \WISE are shown in Fig.\ \ref{fig:SEDclassified}. 
The two unclassified UV-excess sources (UVEXJ2039+3647 and UVEXJ0009+6514) at $(W2-W3)$$\sim$5.5 show a strong excess in the $W3$-band
(UVEXJ0009+6514 has ccd-flag ``H'' in $W2$).
From their SEDs and position in the \WISE colour-colour diagram, these two sources are candidate 
Luminous InfraRed Galaxies (LIRGs)/star-burst Sbc (Fig.\,12 of Wright et al., 2010).
The four remaining sources are DA+dM or sdB/sdO+MS candidates.
\\

\section{Candidate QSOs selected from UVEX and WISE}
\label{sec:QSO}
\begin{figure}
\centerline{\epsfig{file=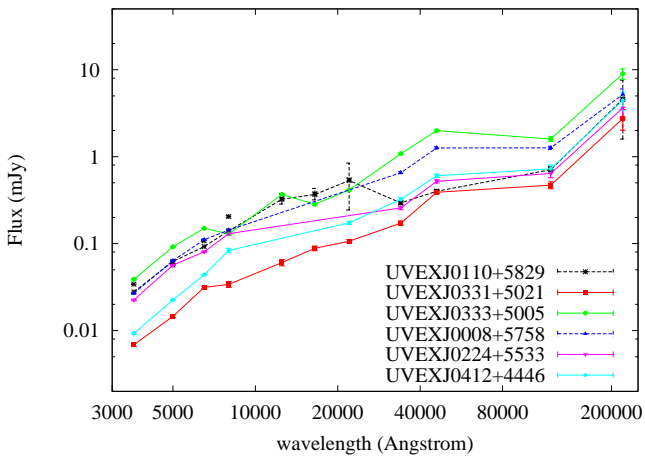,width=9cm,angle=0,clip=}}
\caption{The SEDs of the 2 classified QSOs of V12b (dashed lines) and the 4 new candidate QSOs from WISE (solid lines).
\label{fig:SEDQSO}}
\end{figure}

Since the release of the all-sky \WISE catalog, the data have been used to select Quasi-Stellar 
Objects (QSOs, Bond et al., 2012, Stern et al., 2012, Xue-Bing Wu et al., 2012, Scaringi et al., 2012).
The IR-excess of the QSOs in \WISE is probably due to optically thick material surrounding most QSOs (Roseboom et al., 2012).
In Fig.\ \ref{fig:WISEccd} there are 6 UV-excess sources at the QSO location 
in the \WISE colour-colour diagram (2.5$<$$(W2-W3)$$<$4.5 and 0.6$<$$(W1-W2)$$<$1.7, see Fig.\,12 of Wright et al., 2010).
Two of these sources (UVEXJ0008+5758 and UVEXJ0110+5829) were already spectroscopically classified as QSOs in V12b. 
For UVEXJ0110+5829 $W4$ is an upper limit, the other candidate QSOs have ``good'' photometry in all \WISE bands.
All 6 candidate QSOs have similar \UVEX colours $(g-r)$$\sim$0.8 and $(U-g)$$\sim$--0.2.
In V12a these 6 sources were selected in $g$ vs. $(U-g)$ and not in $g$ vs. $(g-r)$ 
(selection flags ``515'' and ``514'') and they are all at Galactic latitude --5\degr$<$$b$$<$--4\degr
and Galactic longitude 117\degr$>$$l$$>$157\degr. This might be due to the warp and flare of the Milky Way 
at this latitude and longitude (Cabrera-Lavers et al., 2007). 
The effects of the shape are different at different lines of sight, the height of the disk is smaller
at some directions and therefore QSOs may be picked-up by UV-excess surveys.
The two brightest candidate QSOs have a match in 2\MASS, and 3 of the candidate QSOs have a match in \UKIDSS. 
The SEDs of the 2 QSOs and new candidate QSOs are shown in Fig.\ \ref{fig:SEDQSO}.
The characteristics of the 6 UV-excess candidate QSOs are summarized in Table\ \ref{tab:qso}.
\\

\begin{table*}
\caption[]{UV-excess candidate QSOs from WISE \label{tab:qso}}
{\small
\begin{tabular}{ | l | l | l | r | l | l | l | l | r | r | r | r | }
    \hline
Name 			&   $l$ &   $b$ & Field & Selection & ($r$) & ($g$) & ($U$) & $W1$ & $W2$ & $W3$ & $W4$ \\ \hline		 
UVEXJ000848.64+575832.7 & 117.3 & -4.43 &   48 & 515 & 18.625 & 19.494 & 19.402 & 14.189 & 12.837 & 10.016 & 8.126 \\
UVEXJ033309.70+500541.8 & 147.6 & -4.86 & 1243 & 515 & 18.296 & 19.093 & 18.999 & 13.642 & 12.341 &  9.763 & 7.532 \\
UVEXJ033113.88+502156.9 & 147.2 & -4.82 & 1243 & 515 & 19.996 & 21.093 & 20.880 & 15.637 & 14.111 & 11.089 & 8.817 \\
UVEXJ022445.84+553325.9 & 135.9 & -4.94 &  810 & 514 & 18.971 & 19.620 & 19.593 & 15.200 & 13.797 & 10.754 & 8.516 \\
UVEXJ041204.47+444629.6 & 156.1 & -4.80 & 1570 & 515 & 19.631 & 20.623 & 20.552 & 14.954 & 13.637 & 10.618 & 8.298 \\
UVEXJ011037.91+582928.1 & 125.4 & -4.29 &  404 & 515 & 18.821 & 19.518 & 19.360 & 15.056 & 14.083 & 10.643 & 8.262 \\		
    \hline
\end{tabular} \\ 
}
\end{table*}

\section{Discussion and conclusions}
\label{sec:discussion}
There are 46 white dwarfs with an infrared match in \UKIDSS and 3 with a match in 2\MASS.
Seventeen sources in the $(g-r)$$<$0.2 sample turn out to be single white dwarfs. 
These white dwarfs have bright \UVEX $g$-band magnitudes,
have in general a higher $T_{\rm eff}$, and are more reddened compared to the 
candidate DA+dM sources in the UV-excess candidate white dwarf sample.\\

In the $(g-r)$$<$0.2 sample there are 24 DA+dM candidates, which is a fraction of 2$\%$ of the complete 
UV-excess candidate white dwarf sample.
This fraction given here is a lower limit, since the fraction of white dwarfs with a companion is higher 
for the brighter UV-excess sources (see Figs.\ \ref{fig:UVEXccdIR} to\ \ref{fig:UVEXcmdIR}).
If only UV-excess candidate white dwarfs brighter than $g$$<$20 are considered, the fraction of 
white dwarfs with a companion is 4$\%$. If we compare this result to other studies, the DA+dM fraction is
in the range 6-22$\%$ (Farihi, Becklin \& Zuckerman, 2005; Debes, 2011).
The presence of an M-dwarf has a strong influence on the optical colours for the cooler white dwarfs. 
Due to the contribution of the M-dwarf the colours of the most DA+dM sources are redder than $(g-r)$$>$0.2. 
An unreddened DA+M4V has $(g-r)$=0.2 for a $T_{\rm eff}$$\sim$28kK white dwarf 
(Fig.\,1 of Rebassa-Mansergas et al., 2012; Fig.\,2 of Augusteijn et al., 2008), 
while a single DA white dwarf with $T_{\rm eff}$=28kK is $(g-r)$=--0.185.
The effects on the UV and optical spectrum due to the presence of a debris 
disks around white dwarfs is negligible (Zabot et al., 2009).\\
 
There are 7 candidates for white dwarfs with a companion later than type M6V in the UV-excess candidate white dwarf sample. 
These sources are all brighter than $g$$<$20 and the white dwarfs all have an effective temperature lower than $T_{\rm eff}$$<$20kK.
This number is in agreement with the WD+BD fractions of other studies (Farihi, Becklin \& Zuckerman, 2005; Debes, 2011).\\

No convincing debris disk candidates were found. If the expected rate of white dwarfs with a dust/debris disk 
would be $\sim$1$\%$ (Girven et al., 2011), there would be $\sim$4 of them brighter than g$<$20 in the UV-excess catalogue.
Additionally, the IR-excess of one source (UVEXJ0328+5035) can be fully explained by a reddened sdB 
spectrum without the need for a low-mass companion.\\

There are 2 known QSOs and 4 UV-excess candidate QSOs in \WISE. Since the number of QSOs 
from the \WISE cross-matching (6 of 2\,170) is much smaller than the fraction of QSOs found in the spectroscopic 
follow-up of UV-excess sources in V12b (2 of 132), there might be some more QSOs in the UV-excess catalogue.
However, the QSOs at $|b|$$<$5 are expected to be clustered at specific lines of sight where absorption is not so strong.
The fact that all known \UVEX QSOs have a \WISE match shows that adding \WISE data is a good additional selection criteria.
A list of UV-excess candidate QSOs is given in Table\ \ref{tab:qso}. 
The \UVEX colours of these candidate QSOs are at 0.65$<$$(g-r)$$<$1.10 and -0.21$<$$(U-g)$$<$-0.03 
(Figs.\ \ref{fig:UVEXccdIR} and\ \ref{fig:UVEXcmdIR}), which are the colour ranges of known QSOs.\\

\section*{Acknowledgement}
This paper makes use of data collected at the Isaac Newton Telescope, operated on the island of La Palma by the Isaac Newton Group in the
Spanish Observatorio del Roque de los Muchachos of the Inst\'{\i}tuto de Astrof\'{\i}sica de Canarias.
The observations were processed by the Cambridge Astronomy Survey Unit
(CASU) at the Institute of Astronomy, University of Cambridge.
Hectospec observations shown in this paper were obtained at the MMT Observatory, a joint facility of the University of Arizona and the Smithsonian Institution.
We gratefully acknowledge the \IPHAS consortium for making available the IPHAS-IDR data.
The WHT/ISIS spectra were reduced in V12b using IRAF (Image Reduction and Analysis Facility). 
IRAF is distributed by the National Optical Astronomy Observatory, which is operated by the Association of 
Universities for Research in Astronomy (AURA) under cooperative agreement with the National Science Foundation.
The authors would like to thank Detlev Koester for
making available the white dwarf atmosphere model spectra.
The TheoSSA service (http://dc.g-vo.org/theossa), used to retrieve theoretical sdO and sdB spectra for this paper, was constructed 
as part of the activities of the German Astrophysical Virtual Observatory.
This research has made use of the Simbad database and the VizieR catalogue access 
tool, operated at CDS, Strasbourg, France. The original description of the VizieR service was published in A\&AS 143, 23
This publication makes use of data products from the Wide-field Infrared Survey Explorer (\WISE), which is a joint project of the University of California, 
Los Angeles, and the Jet Propulsion Laboratory/California Institute of Technology, funded by the National Aeronautics and Space Administration.
This publication makes use of data products from the Two Micron All Sky Survey (2\MASS), which is a joint project of the University of Massachusetts and the 
Infrared Processing and Analysis Center/California Institute of Technology, funded by the National Aeronautics and Space Administration and the 
National Science Foundation.   
This work is based in part on data obtained as part of the UKIRT Infrared Deep Sky Survey (\UKIDSS).
We want to credit the efforts of the teams which built WFCAM, processed the data, and implemented the \UKIDSS surveys.
The \UKIDSS project is defined in Lawrence et al (2007). \UKIDSS uses the UKIRT Wide Field Camera (WFCAM; Casali et al, 2007). 
The photometric system is described in Hewett et al (2006), and the calibration is described 
in Hodgkin et al. (2009). The pipeline processing and science archive are described in Irwin et al (in prep) and Hambly et al (2008).
This paper uses data obtained by the Sloan Digital Sky Survey (\SDSS) DR8 (Aihara et al. 2011a).
The Sloan Digital Sky Survey III (Eisenstein et al. 2011) is an extension of the SDSS-I
and II projects (York et al. 2000). It uses the dedicated 2.5-meter wide-field Sloan Foundation Telescope (Gunn et al. 2006) at Apache Point Observatory (APO).
Funding for the Sloan Digital Sky Survey has been provided by the Alfred P. Sloan Foundation, 
the Participating Institutions, the National Aeronautics and Space Administration, the National Science 
Foundation, the U.S. Department of Energy, the Japanese Monbukagakusho, and the Max Planck Society. 
The \SDSS Web site is $http://www.sdss.org/$. The \SDSS is managed by the Astrophysical Research Consortium 
(ARC) for the Participating Institutions.
KV is supported by a NWO-EW grant 614.000.601 to PJG and by NOVA.
\\

\label{lastpage}

\appendix

\section{UV-excess sources with an IR-excess}
\label{app:appendix}

\oddsidemargin=-2cm
\evensidemargin=-2cm

\begin{figure*}
\centerline{\epsfig{file=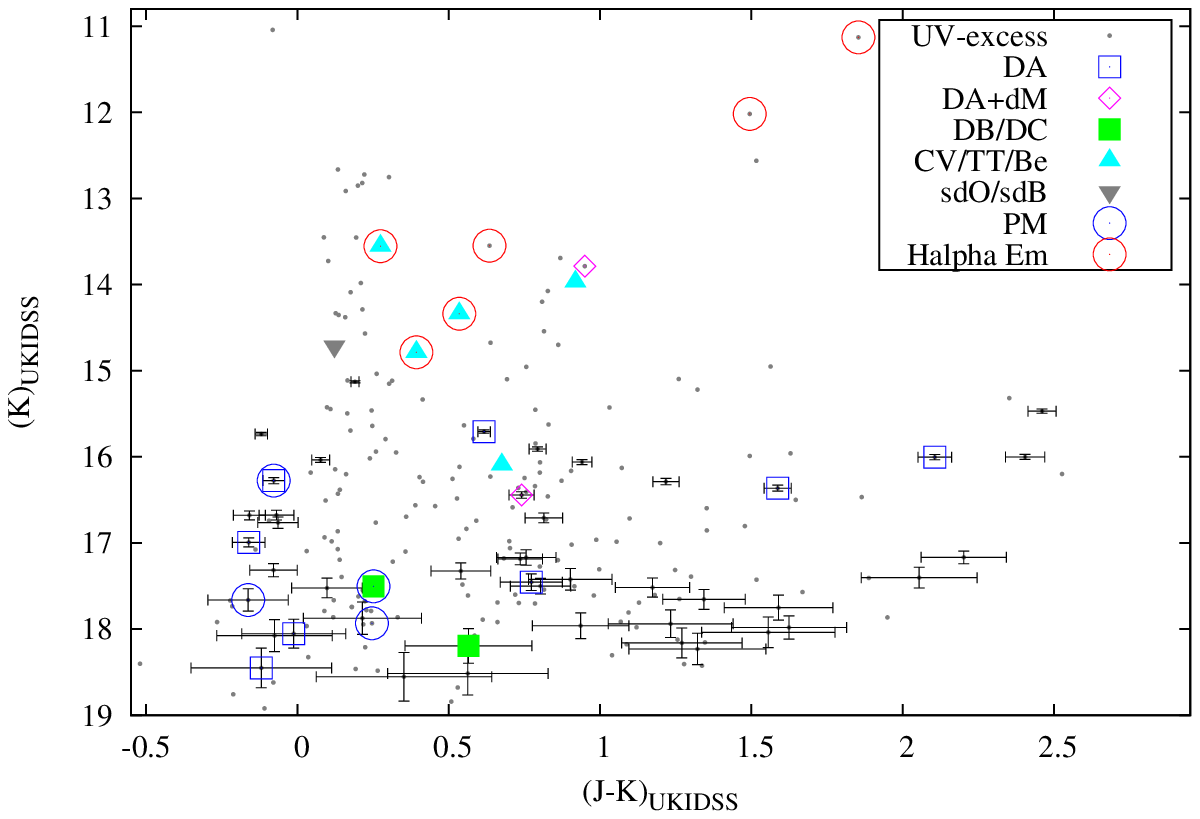,width=14cm,angle=0,clip=}}
\caption{The $K$ vs. $(J-K)$ colour-magnitude diagram with the UKIDSS-GPS matches. 
UV-excess sources spectroscopically classified in V12b are overplotted with different symbols,
UV-excess candidate white dwarfs with $(g-r)<$0.2 are plotted with error bars, 
other UV-excess sources are plotted with dots.
\label{fig:UKIDSScmd}}
\end{figure*}

\begin{figure*}
\centerline{\epsfig{file=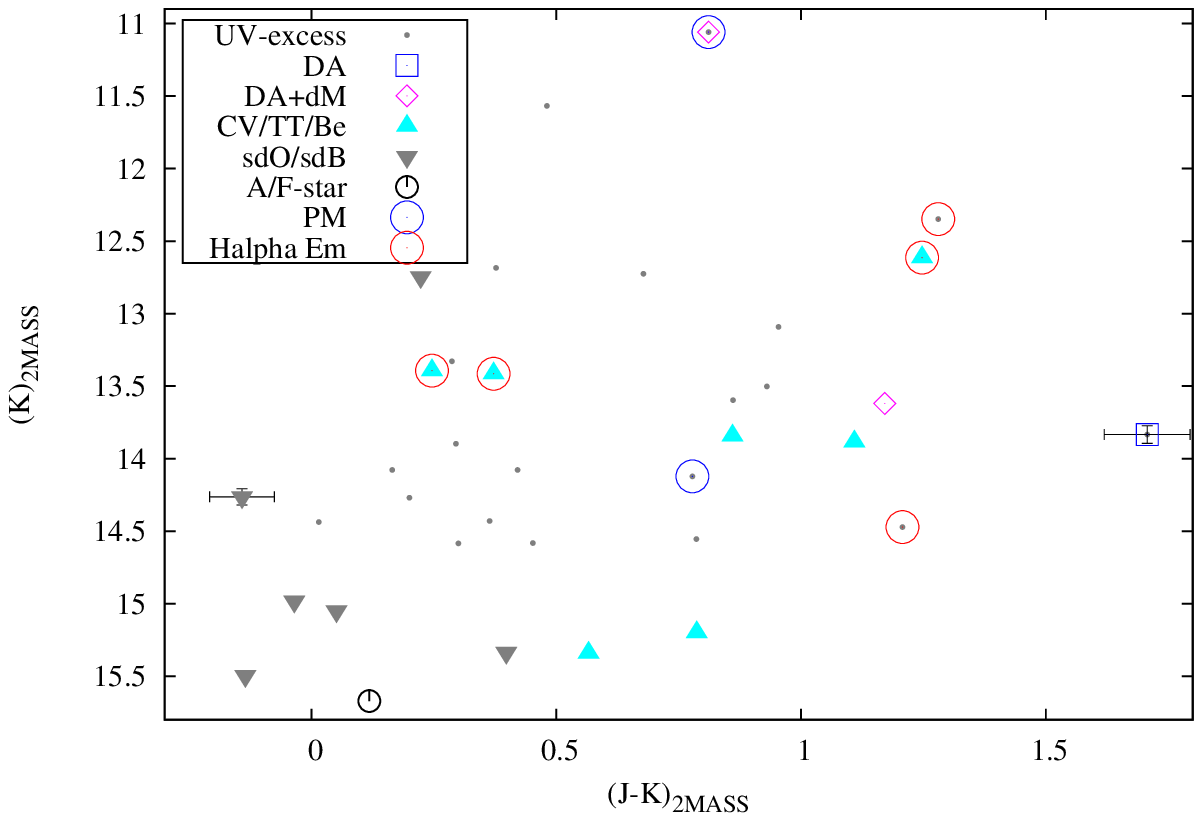,width=14cm,angle=0,clip=}}
\caption{$K$ vs. $(J-K)$ colour-magnitude diagram with the UV-excess matches in 2MASS.
Classified sources are labelled with different symbols, UV-excess candidate white dwarfs are plotted with error bars, 
other UV-excess sources are plotted with dots. There is one more match at $(J-K)$=3.7, $K$=10.8, classified as 
DA white dwarf in V12b, not visible in this figure.
\label{fig:2MASScmd}}
\end{figure*}

\begin{table*}
\caption[]{UV-excess candidate white dwarfs with a match in UKIDSS (46) or 2MASS (3). 
Here ``Selec'' is column 20 of the UV-excess catalogue of V12a, 
``Reg.'' is the region in the $(J-H)$ vs. $(H-K)$ colour-colour diagram,
and ``Fit'' shows the fitting results: i) $T_{\rm eff}$ (kK) of the white dwarf, ii) $E(B-V)$, and iii) the spectral type of companion
(M-type determined by fitting the photometry, or BD determined from the resulting $(J-K)$ colour).\label{tab:IRexcess2}}
\centering
\begin{sideways}
\footnotesize
\noindent
\begin{tabular}{ | l | c | c | c | c | c | c | c | c | c | c | c | c | l | l |   }
    \hline
No  & Name  &  $l$  & $b$ & Field & Selec & $r$ & $g$ & $U$ & $(J)$ & $(H)$ & $(K)$ & Reg. & Fit & V12b \\ \hline	   
1  & UVEXJ183141.67+002201.5 & 31.00431  &  4.55953 &  4088 & 1543 & 17.266 & 17.351 & 16.970 & 17.500 & 17.499 & 17.663 & 1 & 13,0.1 &  \\
2  & UVEXJ183605.84+014117.8 & 32.68689  & 4.18264  & 4112  & 1028 & 18.651 & 18.717 & 17.978 & 17.922 & 17.346 & 17.186 & 2 & 22,0.2,M4V & \\
3  & UVEXJ184610.80+022032.4 & 34.41846  & 2.23713  & 4183  & 1028 & 18.158 & 18.253 & 17.933 & 17.866 & 17.532 & 17.326 & 3 & 14,0.2,L2 & \\
4  & UVEXJ184725.16-011039.4 & 31.42575  &  0.35696 &  4197 & 1543 & 16.992 & 17.208 & 16.804 & 16.700 & 16.592 & 16.764 & 1 & 17,0.3 &  \\
5  & UVEXJ184736.88-004413.1 & 31.84015  & 0.51452  & 4197  & 1543 & 20.435 & 20.594 & 20.508 & 19.457 & 18.016 & 17.403 & 2 & 13,0.3,M4V & \\
6  & UVEXJ185519.35+114741.5 & 43.88889  &  4.49896 &  4294 & 1543 & 18.057 & 18.155 & 17.821 & 18.000 & 18.024 & 18.076 & 1 & 15,0.2 &  \\
7  & UVEXJ185710.76+043917.5 & 37.72853  & 0.84714  & 4338  & 1543 & 19.266 & 19.290 & 18.976 & 17.930 & 16.512 & 15.469 & 2 & 22,0.1,M5V & \\
8  & UVEXJ185941.43+013954.0 & 35.35485  & -1.07596 &  4404 & 1543 & 18.133 & 18.180 & 17.853 & 18.325 & 17.949 & 17.423 & 3 & 13,0.1,L7 & \\
9  & UVEXJ190129.69+075854.1 & 41.17849  &  1.41266 &  4414 & 1543 & 17.800 & 17.815 & 17.046 & 18.089 & 17.842 & 17.874 & 1 & 24,0.2 &  \\
10 & UVEXJ190257.79+113618.9 & 44.56953  & 2.74590  & 4469  & 1028 & 18.495 & 18.508 & 18.009 & 17.524 & 16.889 & 16.710 & 2 & 13,0.0,M6V &  \\
11 & UVEXJ190310.08+140658.9 & 46.83175  &  3.84480 &  4453 & 1028 & 17.595 & 17.667 & 16.743 & 17.621 & 17.578 & 17.524 & 1 & 35,0.3 &   \\
12 & UVEXJ191001.10+055542.5 & 40.32597  & -1.40915 &  4591 & 1543 & 18.154 & 18.209 & 17.585 & 17.926 & 17.526 & 17.170 & 3 & 20,0.2,L5 &  \\
13 & UVEXJ202432.88+412338.9 & 79.33309  &  2.15868 &  5889 & 1543 & 17.351 & 17.207 & 16.286 & 17.835 & 18.387 & 19.690 & 1 & 22,0.0 &   \\
14 & UVEXJ202654.69+430152.1 & 80.92379  & 2.74656  & 5923  & 1028 & 21.309 & 21.491 & 21.140 & 19.594 & 18.617 & 18.038 & 2 & 11,0.1,M6V &  \\
15 & UVEXJ202659.21+411644.1 & 79.50370  & 1.71848  & 5916  & 1028 & 17.175 & 17.092 & 16.389 & 16.323 & 15.996 & 15.707 & 3 & 17,0.0,L3 &  DA \\
16 & UVEXJ202701.05+405909.2 & 79.26842  & 1.54359  & 5916  & 1028 & 20.996 & 21.067 & 20.797 & 19.607 & 18.666 & 17.982 & 2 & 14,0.2,M2V &  \\
17 & UVEXJ202800.47+405620.0 & 79.33903  &  1.36420 &  5939 & 1028 & 16.328 & 16.177 & 17.676 & 16.833 & 16.898 & 16.994 & 1 & 14,0.1 &  DA \\
18 & UVEXJ202940.45+424613.7 & 81.00707  & 2.18354  & 5947  & 1028 & 19.839 & 19.939 & 19.418 & 17.505 & 16.758 & 16.287 & 2 & 50,0.2,M2V &  \\
19 & UVEXJ203238.52+411339.4 & 80.08644  & 0.82801  & 6010  & 1543 & 19.446 & 19.295 & 18.520 & 18.110 & 16.700 & 16.004 & 2 & 50,0.0,M6V &  DA \\
20 & UVEXJ203326.92+410959.9 & 80.12765  & 0.66975  & 6010  & 1543 & 20.967 & 21.135 & 20.775 & 18.406 & 16.892 & 16.001 & 2 & 13,0.1,M6V &  \\
21 & UVEXJ203614.30+392309.8 & 79.02005  & -0.82249 &  6036 & 1029 & 19.582 & 19.776 & 19.192 & 17.184 & 16.662 & 16.443 & 2 & 45,0.4,M2V &  DA+dM \\
22 & UVEXJ204154.36+393201.4 & 79.80276  & -1.60107 &  6118 & 1543 & 19.196 & 19.360 & 19.019 & 18.306 & 17.646 & 17.503 & 2 & 14,0.2,M4V &  \\
23 & UVEXJ204203.59+413343.1 & 81.42044  & -0.37688 &  6112 & 1543 & 21.044 & 21.199 & 20.485 & 19.370 & 17.779 & 17.168 & 2 & 30,0.4,M0V &  \\
24 & UVEXJ204212.01+402706.7 & 80.56161  & -1.08086 &  6111 & 1543 & 18.828 & 19.018 & 18.172 & 16.703 & 16.179 & 15.910 & 2 & 75,0.5,M6V &  \\
25 & UVEXJ204229.67+384058.0 & 79.20167  & -2.21444 &  6108 & 1543 & 19.194 & 19.344 & 18.904 & 19.079 & 18.707 & 18.516 & 3 & 16,0.2,L2 &  \\
26 & UVEXJ204401.03+403014.5 & 80.81605  & -1.32043 &  6143 & 1543 & 15.436 & 15.419 & 15.044 & 15.616 & 15.667 & 15.736 & 1 & 14,0.1 &   \\
27 & UVEXJ204502.52+414622.9 & 81.93084  & -0.68385 &  6145 & 1543 & 17.255 & 17.406 & 16.741 & 17.238 & 17.276 & 17.317 & 1 & 28,0.4 &   \\
28 & UVEXJ204502.62+385449.5 & 79.69073  & -2.46193 &  6172 & 1028 & 21.678 & 21.666 & 21.370 & 19.553 & 18.449 & 18.231 & 2 & 30,0.1,M4V &  \\
29 & UVEXJ204503.63+435657.2 & 83.63666  & 0.66749  & 6153  & 1028 & 19.234 & 19.171 & 18.334 & 18.691 & 18.013 & 17.518 & 2 & 75,0.1,M5V &  \\
30 & UVEXJ204628.29+433237.6 & 83.47876  & 0.21540  & 6176  & 1543 & 20.841 & 20.827 & 20.039 & 18.999 & 18.078 & 17.656 & 2 & 20,0.1,M2V &  \\
31 & UVEXJ204720.64+444310.3 & 84.49330  & 0.82997  & 6196  & 1543 & 18.436 & 18.572 & 18.181 & 17.001 & 16.279 & 16.061 & 2 & 15,0.2,M2V &  \\
32 & UVEXJ204751.27+442920.1 & 84.37091  & 0.61442  & 6196  & 1028 & 20.004 & 19.960 & 19.457 & 17.952 & 16.947 & 16.365 & 2 & 14,0.0,M6V &  DA \\
33 & UVEXJ204804.01+421720.8 & 82.68492  & -0.79916 &  6209 & 1543 & 19.721 & 19.677 & 19.124 & 19.431 & 18.456 & 18.161 & 2 & 14,0.0,M6V &  \\
34 & UVEXJ204830.05+423250.4 & 82.93598  & -0.69880 &  6188 & 1031 & 21.198 & 21.121 & 20.907 & 19.172 & 18.430 & 17.939 & 2 & 13,0.1,M0V &  \\
35 & UVEXJ204856.21+444455.0 & 84.69394  & 0.62843  & 6196  & 1028 & 18.638 & 18.513 & 17.832 & 18.898 & 18.502 & 17.962 & 3 & 16,0.0,L7 &  \\
36 & UVEXJ204914.51+421623.0 & 82.80950  & -0.97798 &  6209 & 1028 & 16.610 & 16.784 & 16.434 & 16.608 & 16.554 & 16.677 & 1 & 14,0.2 &   \\
37 & UVEXJ204938.33+432018.0 & 83.68083  & -0.36073 &  6199 & 1543 & 19.137 & 19.328 & 18.962 & 18.906 & 18.721 & 18.555 & 3 & 11,0.1,M8V &  \\
38 & UVEXJ205037.81+424618.9 & 83.35761  & -0.85988 &  6220 & 1543 & 15.793 & 15.750 & 15.132 & 16.198 & 16.242 & 16.277 & 1 & 15,0.0 &  DA \\
39 & UVEXJ205148.13+442408.8 & 84.75023  &  0.01441 &  6219 & 1028 & 17.806 & 17.707 & 16.905 & 18.333 & 18.499 & 18.452 & 1 & 18,0.0 &  DA \\
40 & UVEXJ205636.98+434541.6 & 84.81618  & -1.05975 &  6293 & 1028 & 19.694 & 19.666 & 19.196 & 19.342 & 18.285 & 17.752 & 2 & 13,0.0,M6V &  \\
41 & UVEXJ210112.02+452020.7 & 86.54346  & -0.64434 &  6355 &  519 & 16.131 & 16.272 & 17.069 & 16.114 & 16.056 & 16.037 & 1 & 14,0.8 &   \\
42 & UVEXJ210248.44+475058.9 & 88.60784  & 0.81096  & 6341  & 1543 & 18.313 & 18.300 & 17.879 & 18.228 & 17.726 & 17.456 & 2 & 80,0.1,L5 &  DA \\
43 & UVEXJ210454.41+460041.9 & 87.47634  & -0.68123 &  6365 & 1543 & 18.778 & 18.738 & 17.856 & 18.762 & 18.334 & 18.196 & 1 & 24,0.1 &  DB \\
44 & UVEXJ223634.77+591907.8 & 106.47749 &  0.82554 &  7139 & 1028 & 17.752 & 17.675 & 16.937 & 18.041 & 18.032 & 18.054 & 1 & 17,0.0 &  DA \\
45 & UVEXJ224338.89+550318.5 & 105.25732 & -3.37032 &  7188 &  519 & 16.501 & 16.547 & 15.781 & 16.521 & 16.546 & 16.680 & 1 & 24,0.2 &   \\
46 & UVEXJ224436.07+544812.6 & 105.25981 & -3.65677 &  7188 &  518 & 15.640 & 15.780 & 15.359 & 15.319 & 15.163 & 15.129 & 1 & 16,0.2 &   \\
 \hline
47 & UVEXJ032855.25+503529.8 & 146.76969 & -4.84547 & 1224  &  519 & 14.078 & 14.205 & 13.591 & 14.121 & 14.091 & 14.263 & 1 & sdO/sdB & sdB \\
48 & UVEXJ185740.07+075557.3 & 40.70227  & 2.23345  & 4346  & 1028 & 18.821 & 19.024 & 18.656 & 15.540 & 14.402 & 13.833 & 2 & DA+M/L5 & DA \\
49 & UVEXJ203411.72+411020.3 & 80.21601  & 0.56034  & 6010  & 1543 & 20.446 & 20.426 & 19.875 & 14.512 & 11.915 & 10.767 & 2 & DA+M/L8 & DA \\
    \hline 
\end{tabular}									       
\end{sideways}
\end{table*}


\label{lastpage}


\begin{thebibliography}{99}
\bibitem[\protect\citeauthoryear{Adelman-McCarthy et al.}{2011}]{adelman-mccarthy11}
Adelman-McCarthy, J. K.; et al., 2011, yCat 2306, 0A
\bibitem[\protect\citeauthoryear{Aihara et al.}{2011}]{aihara11}
Aihara H., Allende Prieto C., An D., Anderson S.F. Aubourg E., Balbinot E., Beers T.C., et al., 2011, ApJS 193, 29A
\bibitem[\protect\citeauthoryear{Augusteijn et al.}{2008}]{augusteijn08}
Augusteijn T., Greimel R., van den Besselaar E.J.M., Groot P.J., Morales-Rueda L., 2008, A\&A 486, 843A
\bibitem[\protect\citeauthoryear{Barber et al}{2012}]{barber12}
Barber S.D., Patterson A.J., Kilic L., Leggett S.K., Dufour P., Bloom J.S., Starr D.L., 2012, ApJ 760, 26
\bibitem[\protect\citeauthoryear{Barentsen et al.}{2011}]{barentsen11}
Barentsen G., Vink J.S., Drew J.E., Greimel R. et al., 2011, MNRAS 415, 103B
\bibitem[\protect\citeauthoryear{Beuermann}{2006}]{beuermann06}
Beuermann K, 2006, A\&A 460, 783B
\bibitem[\protect\citeauthoryear{Blanton \& Roweis}{2007}]{blanton07}
Blanton M.R. and Roweis S., 2007, AJ 133, 734B	
\bibitem[\protect\citeauthoryear{Brinkworth et al}{2009}]{brinkworth09}
Brinkworth C.S., G{\"a}nsicke B.T., Marsh T.R., Hoard D.W., Tappert C., 2009, ApJ 696, 1402B
\bibitem[\protect\citeauthoryear{Brinkworth et al.}{2012}]{brinkworth12}
Brinkworth C.S., G{\"a}nsicke B.T., Girven J.M., Hoard D.W., Marsh T.R., Parsons S.G., Koester D., 2012, ApJ 750, 86B
\bibitem[\protect\citeauthoryear{Cabrera-Lavers et al.}{2007}]{cabrera-Lavers07}
Cabrera-Lavers A., Bilir S., Ak S., Yaz E., Lopez-Corredoira M., 2007, A\&A 464, 565C
\bibitem[\protect\citeauthoryear{Cardelli et al.}{1989}]{ccm89}
Cardelli J.A., Clayton G.C. \& Mathis J.S., 1989, ApJ 345, 245
\bibitem[\protect\citeauthoryear{Casali et al.}{2007}]{casali07}
Casali M., Adamson A., Alves de Oliveira C., Almaini O., Burch K., Chuter T., Elliot J. et al., 2007, A\&A 467, 777C
\bibitem[\protect\citeauthoryear{Corradi et al.}{2010}]{corradi10}
Corradi R. L. M., Valentini M., Munari U., Drew J. E., et al., 2010, A\&A 509, 41
\bibitem[\protect\citeauthoryear{Corradi et al.}{2011}]{corradi11}
Corradi R.L.M., Sabin L., Miszalski B., Rodríguez-Gil P., Santander-García M., Jones D., Drew J.E. et al., 2011, MNRAS 410, 1349C
\bibitem[\protect\citeauthoryear{Cutri et al.}{2003}]{cutri03}
Cutri R.M., Skrutskie M.F., van Dyk S. et al., 2003, yCat 2246, 0C
\bibitem[\protect\citeauthoryear{Cutri et al.}{2012}]{cutri12}
Cutri R.M. et al., 2012, yCat 2311, 0C
\bibitem[\protect\citeauthoryear{Deacon et al.}{2009}]{deacon09}
Deacon N.R., Groot P.J., Drew J.E., et al., 2009, MNRAS.397, 1685
\bibitem[\protect\citeauthoryear{Debes et al.}{2011}]{debes11}
Debes J.H., Hoard D.W., Wachter S., Leisawitz D.T., Cohen M., 2011, ApJS 197, 38D
\bibitem[\protect\citeauthoryear{Debes et al}{2012}]{debes12}
Debes J.H., Walsh K.J., Stark C., 2012, ApJ 747, 148D
\bibitem[\protect\citeauthoryear{Debes et al.}{2011}]{debes11}
Debes J.H., Hoard D.W., Kilic M., Wachter S., Leisawitz D.T., Cohen M., Kirkpatrick J.D., Griffith R.L., 2011, ApJ 729, 4D
\bibitem[\protect\citeauthoryear{Debes \& Sigurdsson}{2002}]{debessigurdsson02}
Debes J.H. \& Sigurdsson, S., 2002, ApJ 572, 556D
\bibitem[\protect\citeauthoryear{Drew et al.}{2005}]{drew05}
Drew J.E., Greimel R., Irwin M., et al., 2005, MNRAS 362, 753 (D05)
\bibitem[\protect\citeauthoryear{Dufour et al}{2012}]{dufour12}
Dufour P., Kilic M., Fontaine G., Bergeron P., Melis C., Bochanski J., 2012, ApJ 749, 6D
\bibitem[\protect\citeauthoryear{Eisenstein et al.}{2011}]{eisenstein11}
Eisenstein D.J., Weinberg D.H., Agol E., Aihara H., Allende Prieto C., Anderson S.F., Arns J.A., Aubourg, E., et al., 2011, AJ 142, 72E
\bibitem[\protect\citeauthoryear{Farihi et al}{2012}]{farihi12}
Farihi J., G{\"a}nsicke B.T., Steele P.R., Girven J., Burleigh M.R., Breedt E., Koester D., 2012, MNRAS 421, 1635F	
\bibitem[\protect\citeauthoryear{Farihi}{2005}]{fahiri05}
Farihi J., Becklin E.E. \& Zuckerman B., 2005, ApJS 161, 394F
\bibitem[\protect\citeauthoryear{G{\"a}nsicke et al.}{2008}]{gansicke08}
G{\"a}nsicke B.T., Koester D., Marsh T.R., Rebassa-Mansergas A., Southworth J., 2008, MNRAS 391L, 103G
\bibitem[\protect\citeauthoryear{G{\"a}nsicke}{2011}]{gansicke11}
G{\"a}nsicke, Boris T., 2011, AIPC 1331, 211G
\bibitem[\protect\citeauthoryear{G{\"a}nsicke et al}{2007}]{gansicke07}
G{\"a}nsicke B.T., Marsh T.R., Southworth J., 2007, MNRAS 380L, 35G
\bibitem[\protect\citeauthoryear{G{\"a}nsicke et al}{2012}]{gansicke12}
G{\"a}nsicke B.T., Koester D., Farihi J., Girven J., Parsons S.G., Breedt E., 2012, MNRAS 424, 333G
\bibitem[\protect\citeauthoryear{Girven et al.}{2011}]{girven11}
Girven J., G{\"a}nsicke B.T., Steeghs D. \& Koester D., 2011, MNRAS 417, 1210G
\bibitem[\protect\citeauthoryear{Gonz\'{a}lez-Solares et al.}{2008}]{gonzalez2008}
Gonz\'{a}lez-Solares E.A., Walton N.A., Greimel R., Drew, J.E., et al., 2008, MNRAS 388, 89 
\bibitem[\protect\citeauthoryear{Greiss et al.}{2012}]{greiss12}
Greiss S., Steeghs D., G{\"a}nsicke B.T., Martín E.L., Groot P.J. et al., 2012 AJ 144, 24G
\bibitem[\protect\citeauthoryear{Groot et al.}{2009}]{groot09}
Groot P.J., Verbeek K., Greimel R., et al., 2009, MNRAS 399, 323G
\bibitem[\protect\citeauthoryear{Gunn et al.}{2006}]{gunn06}
Gunn J.E., Siegmund W.A., Mannery E.J., Owen R.E., Hull C.L., Leger R.F, Carey L.N. et al., 2006, AJ 131, 2332G
\bibitem[\protect\citeauthoryear{Hales et al.}{2009}]{hales09}
Hales, A.S., Barlow M.J., Drew J.E., Unruh Y.C., Greimel R., Irwin M.J., González-Solares E., 2009, ApJ 695, 75H
\bibitem[\protect\citeauthoryear{Hambly et al.}{2008}]{hambly08}
Hambly N.C., Collins R.S., Cross N.J.G., Mann R.G., Read M.A., Sutorius E.T.W., Bond I., Bryant J., et al., 2008, MNRAS 384, 637H
\bibitem[\protect\citeauthoryear{Heller et al}{2009}]{heller09}
Heller R., Homeier D., Dreizler S., Oestensen R., 2009, yCat 34960191H
\bibitem[\protect\citeauthoryear{Hewett et al.}{2006}]{hewett06}
Hewett P.C., Warren S.J., Leggett S.K., Hodgkin S.T., 2006, MNRAS 367, 454H
\bibitem[\protect\citeauthoryear{Hoard et al}{2011}]{hoard11}
Hoard D.W., Debes J.H., Wachter S., Leisawitz D.T., Cohen M., 2011, AAS 21733309H
\bibitem[\protect\citeauthoryear{Hodgkin et al.}{2009}]{hodgkin09}	
Hodgkin S.T., Irwin M.J., Hewett P.C. \& Warren S.J., 2009, MNRAS 394, 675H
\bibitem[\protect\citeauthoryear{Jura}{2003}]{jura03}
Jura M., 2003, ApJ 584L, 91J
\bibitem[\protect\citeauthoryear{Kilic et al.}{2012}]{kilic12}
Kilic M., Patterson A.J., Barber S., Leggett S.K., Dufour P., 2012, MNRAS 419L, 59K
\bibitem[\protect\citeauthoryear{Koester et al.}{2001}]{koester01}
Koester D., et al., 2001, A\&A, 378, 556
\bibitem[\protect\citeauthoryear{Koester}{2009}]{koester09}
Koester D., 2009, A\&A 498,517K
\bibitem[\protect\citeauthoryear{Koester \& Wilken}{2006}]{koesterwilken06}
Koester D. \& Wilken D., 2006, A\&A 453, 1051K
\bibitem[\protect\citeauthoryear{Lawrence et al.}{2007}]{lawrence07}
Lawrence A., Warren S.J., Almaini O., Edge A.C., Hambly N.C., Jameson R.F., Lucas P., Casali M., et al., 2007, MNRAS 379, 1599L
\bibitem[\protect\citeauthoryear{Lawrence et al.}{2012}]{lawrence12}
Lawrence A., Warren S.J., Almaini O., Edge A.C., Hambly N.C., Jameson R.F., Lucas P., Casali M., Adamson A., et al., 2012, yCat 2314, 0L 
\bibitem[\protect\citeauthoryear{Leggett et al}{2002}]{leggett02}
Leggett S.K., Golimowski D.A., Fan X., Geballe T.R., Knapp G.R., Brinkmann J. et al., 2002, ApJ 564, 452L
\bibitem[\protect\citeauthoryear{Lucas et al.}{2008}]{lucas08}
Lucas P. W., et al, 2008, 2008MNRAS.391..136L
\bibitem[\protect\citeauthoryear{Pickles A.J.}{1998}]{pickles98}
Pickles A.J., 1998, PASP 110, 863 
\bibitem[\protect\citeauthoryear{Rebassa-Mansergas et al}{2010}]{rebassa-mansergas10}
Rebassa-Mansergas A., G{\"a}nsicke B.T., Schreiber M.R., Koester D. and Rodr\'iguez-Gil P., 2010, MNRAS 402, 620
\bibitem[\protect\citeauthoryear{Rebassa-Mansergas et al.}{2012}]{rebassa-mansergas12}
Rebassa-Mansergas A., Nebot Gómez-Morán A., Schreiber M.R., G{\"a}nsicke B.T., Schwope A., Gallardo J., Koester D., 2012, MNRAS 419, 806R
\bibitem[\protect\citeauthoryear{Reid et al}{2001}]{reid01}
Reid I.N., Burgasser A.J., Cruz K.L., Kirkpatrick J.D., Gizis J.E., 2001, AJ 121, 1710R
\bibitem[\protect\citeauthoryear{Ringat}{2012}]{ringat12}
Ringat E., 2012, ASPC 452, 99R
\bibitem[\protect\citeauthoryear{Roseboom et al}{2012}]{roseboom12}
Roseboom I.G., Lawrence A., Elvis M., Petty S., Shen Y., Hao H., 2012, arXiv 1205, 4543v2
\bibitem[\protect\citeauthoryear{Scaringi et al.,}{2013}]{scaringi13}
Scaringi S., Groot P.J., Verbeek K., Greiss S., Knigge C. \& Koerding E., 2013, MNRAS 428, 2207S  
\bibitem[\protect\citeauthoryear{Silvestri et al}{2006}]{silvestri06}
Silvestri N.M., Hawley S.L., West A.A., Szkody P., Bochanski J.J., Eisenstein D.J. et al., 2006, AJ 131, 1674S
\bibitem[\protect\citeauthoryear{Skrutskie et al.,}{2006}]{skrutskie06}
Skrutskie M.F., Cutri R.M., Stiening R., Weinberg M.D., Schneider S., Carpenter J.M., Beichman C., Capps R., et al., 2006, AJ 131, 1163S
\bibitem[\protect\citeauthoryear{Steele et al.}{2011}]{steele11}
Steele P.R., Burleigh M.R., Dobbie P.D., Jameson R.F., Barstow M.A., Satterthwaite R.P., 2011, MNRAS 416, 2768S
\bibitem[\protect\citeauthoryear{Stern et al.}{2012}]{stern12}
Stern, D., Assef R.J., Benford D.J., Blain A., Cutri R., et al., 2012, ApJ 753, 30S
\bibitem[\protect\citeauthoryear{Verbeek et al.}{2012}]{verbeek12a}
Verbeek K., Groot P.J., de Groot E., Scaringi S., Drew J.E., et al., 2012, MNRAS 420, 1115V
\bibitem[\protect\citeauthoryear{Verbeek et al.}{2012}]{Verbeek12b}	
Verbeek K., Groot P.J., Scaringi S., Napiwotzki R., Spikings B., {\O}stensen R.H., Drew J.E., Steeghs D. et al, 2012, MNRAS 426, 1235V
\bibitem[\protect\citeauthoryear{Wachter et al.}{2003}]{wachter03}
Wachter S., Hoard D.W., Hansen K.H., Wilcox R.E., Taylor H.M. \& Finkelstein S.L., 2003, ApJ 586, 1356W
\bibitem[\protect\citeauthoryear{Wesson et al.}{2008}]{wesson08}
Wesson R., Barlow M., Corradi R., et al., 2008, ApJ 688, L21
\bibitem[\protect\citeauthoryear{Witham et al.}{2008}]{witham08}
Witham A.R., Knigge C., Drew J.E., et al., 2008, MNRAS 384, 1277 
\bibitem[\protect\citeauthoryear{Wright et al}{2008}]{wright08}
Wright N.J., Greimel R., Barlow M.J., Drew J.E., Cioni M.R.L., Zijlstra A.A. et al., 2008, MNRAS 390, 929W
\bibitem[\protect\citeauthoryear{Wright et al}{2012}]{wright12}
Wright N.J., Drake J.J., Drew J.E. et al., 2012, ApJ 746L, 21W
\bibitem[\protect\citeauthoryear{Wu et al.}{2012}]{wu99}
Wu X.B., Hao G., Jia Z., Zhang Y. \& Peng N., 2012, AJ 144, 49W
\bibitem[\protect\citeauthoryear{York et al}{2000}]{york00}
York D.G., Adelman J., Anderson J.E. et al., 2000, AJ 120,1579Y
\bibitem[\protect\citeauthoryear{Zabot et al}{2009}]{zabot09}
Zabot A., Kanaan A., Cid Fernandes R., 2009, ApJ 704L, 93Z
\bibitem[\protect\citeauthoryear{Zuckerman and Becklin}{1987}]{zuckerman87}
Zuckerman B., \& Becklin E E., 1987, Natur 330, 138Z
\end{thebibliography}
\end{document}